\documentclass[12pt]{iopart}
\usepackage{iopams}
\usepackage{setstack}
\usepackage{cite}
\usepackage{cleveref}
\usepackage{graphicx}
\usepackage{array,multirow}
\usepackage[T1]{fontenc}
\usepackage{float}
\usepackage{dcolumn}
\usepackage{xcolor}

\newcommand{\beq}{\begin{equation}}
\newcommand{\eeq}{\end{equation}}
\newcommand{\bea}{\begin{eqnarray}}
\newcommand{\eea}{\end{eqnarray}}

\begin{document}

\title{Mode entanglement and isospin pairing in two-nucleon systems}

\author{J Kov\'acs$^1$\footnote{Author to whom any correspondence should be addressed.},  A T Kruppa$^1$, {\" O} Legeza$^{2,3,4}$, P Salamon$^{1}$}

\address{$^1$HUN-REN Institute for Nuclear Research, Bem tér 18/c, Debrecen H-4026, Hungary}
\address{$^2$HUN-REN Wigner Research Centre for Physics, Konkoly-Thege Miklós út 29-33, Budapest H-1121, Hungary}
\address{$^3$Institute for Advanced Study, Technical University of Munich, Germany, Lichtenbergstrasse 2a, 85748 Garching, Germany}
\address{$^4$Parmenides Stiftung, Hindenburgstr. 15, 82343, Pöcking, Germany}   
\eads{ \mailto{kovacs.jozsef@atomki.hun-ren.hu}, \mailto{atk@atomki.hun-ren.hu}, \mailto{legeza.ors@wigner.hun-ren.hu}, 
 \mailto{salamon@atomki.hun-ren.hu}}

\begin{abstract}

In this study, we explore the entanglement and correlation in two-nucleon systems using isospin formalism. With the help of  Slater decomposition, we derive analytical expressions for various entanglement measures. Specifically, we analyse the one- and two-mode entropies, mutual informations, and a basis-independent characteristic known as the one-body entanglement entropy.

To understand the impact of pairing, we consider interactions involving isovector and isoscalar $L=0$ pairing terms. Our findings show that certain pairing interactions can maximize one-body entanglement entropy of ground states when both total angular momentum and total isospin have zero projections.

We provide numerical examples for the sd shell and explore the mutual informations in $LS$ coupled and 
$jj$ coupled single-particle bases. We find that the shell structure and angular momentum coupling significantly impact the measures of entanglement. 
We outline the implications of conserving angular momentum and isospin on one-mode entropies, irrespective of particle number.

\end{abstract}
\submitto{\jpg}

\maketitle
\section{Introduction}

Entanglement is a fundamental concept in quantum mechanics, highlighting the correlations between particles or partitions within a system that cannot be independently described. Significant effort has been devoted to studying entanglement in various areas of physics, including its role in many-body problems \cite{ami08} and systems of indistinguishable particles \cite{Benatti20,szi21}. Possible applications of the entanglement for basic science and technology are reviewed in \cite{klc22}.

Entanglement is based on the tensor product of the subsystems' Hilbert spaces for distinguishable particles, as thoroughly investigated \cite{hor09}. Fermions are described by the antisymmetric part of the Fock space, but the decomposition into particle subsystems does not align with the Fock space's tensor product structure.
Various approaches address the definition of subsystems for fermions. The mode-entanglement method \cite{zan02,leg03,shi03,git02}, 
uses the second quantized formalism where,
the mode creation and annihilation operators generate the algebra of observables. Subsystems
are defined by subalgebras \cite{ben14,ben14b,ban07}.
The algebraic approach is based on correlations between observables \cite{ben16}, and descriptions relying on quantum correlations of particles is also considered \cite{eck02,pas01,sch01,Ghirardi02,din20}.
Several suggested approaches are reviewed in \cite{Benatti20}.

Our understanding of entanglement in nuclei is still in its early stages, but progress has been made in studying entanglement in nuclear physics models. These studies include investigations correlations in the Lipkin model and its extensions \cite{lat05,tul19,fab21,Faba22}, as well as in fermionic superconducting systems \cite{tul18}.
Within the traditional nuclear shell model framework, researchers have explored the entanglement of particles \cite{kwa14,kwa16,kwa17}, mode entanglement \cite{leg15,Kruppa21,Kruppa22,Stetcu22,Tichai23,tic24} and recently multi-partite entanglement is also investigated \cite{bro24}. 

Using realistic effective interactions and the traditional shell model, selected nuclei in the p, sd and pf shells have been studied \cite{PerezObiol23a} and general relationships between some traditional nuclear physics concepts and measures of entanglement are established.
The paper  \cite{sav} explores the entanglement patterns resulting from no-core shell model calculations of light helium nuclei. It demonstrates how specific transformations of the effective Hamiltonian can speed up the convergence of the wave function and binding energy while reducing and relocating entanglement within the effective model space. 
In the framework of the coupled-cluster method  entanglement based on a
partition of the single-particle space into holes and particles are also studied in \cite{Gu23}.
More recently, entanglement has been investigated in algebraic models \cite{Jafarizadeh21,Jafarizadeh24} to describe phase transitions and in the context of nuclear scattering \cite{Bai22,Bai23a,Bai24b,Miller23a,Miller23b,li24}.

In complex nuclear systems, exact calculations are nearly impossible. Therefore, effective methods, such as truncating the Hilbert space and transforming the Hamiltonian using a unitary transformation, are used. Entanglement provides measures and insight that can help develop better effective descriptions of nuclear systems. A study on this topic in the Lipkin-Meshkov-Glick model was published 
\cite{Illa23,Hengstenberg23,Robin23}.

The entanglement of protons and neutrons in both ground and excited states has been extensively studied in the sd shell \cite{Johnson23}. It has been found that nuclear shell model Hamiltonians show low entanglement between the proton and neutron components of the wave function. This finding helps to explain the effectiveness of the proton-neutron singular value decomposition method of shell model wave functions \cite{pap03, pap04}. 
By utilizing low entanglement between protons and neutrons, a new approximation method has been developed that significantly reduces the basis size of shell model calculations \cite{gor24}.

Two formalisms can be used to describe atomic nuclei: isospin and proton-neutron formalism. The practical choice between these approaches depends on the specific context of the problem at hand.
The isospin formalism treats protons and neutrons as two states of a single-particle, emphasizing their similarities under the strong interaction. This unified perspective can enhance our understanding of nuclear phenomena, as isospin conservation is often a valid approximation in nuclear systems.
In the isospin formalism, the wave function must adhere to the generalized exclusion principle, which requires antisymmetry with respect to all coordinates: spatial, spin, and isospin. 

In our current work, we utilize the isospin formalism, treating the particles as indistinguishable.
This formalism offers a straightforward way to investigate mode entanglement in second quantization formalism. 
It would be beneficial to explore the proton-neutron formalism by treating the proton-neutron system as distinguishable fermions and examining when it behaves like an elementary boson, as outlined in \cite{chu10}.

We examine the implications of rotational and isospin symmetries of nuclear wave functions, as these symmetries are prevalent in nuclear structure models. Understanding these effects can help differentiate between interaction-specific outcomes and symmetry consequences, aiding the analysis of complex models.
Our focus is on angular momentum and isospin coupled two-nucleon states, providing analytical and numerical results for one-body entanglement entropies and mutual informations \cite{ris06,din21}. 

In our previous paper \cite{Kruppa21} proton-proton and neutron-neutron two-body systems were studied. In isospin formalism they correspond to states with isospin quantum numbers $T=1,\ T_z=1$ and $T=1,\ T_z=-1$, respectively. Now, we extend the investigation to two-body states with quantum numbers $T=0,1$ and $T_z=0$, corresponding to proton-neutron systems.
Discussing the entanglement of two-nucleon systems using the isospin formalism, we have not only consider the new degree of freedom but also investigate the consequences of isospin symmetry. We derive new analytical formulas for wave functions with good angular momentum and total isospin. We examine the implications of isospin symmetry for one-mode entropies for arbitrary number of particles. We prove that when the number of protons equals the number of neutrons, the one-mode entropies for protons and neutrons are identical in wave functions with good isospin. This insight may serve as a valuable tool for quantifying isospin breaking.

Two-fermion systems provide a natural starting point to investigate entanglement. They offer significant technical advantages since they can be transformed into what is known as Slater decomposed form \cite{pas01,sch01}. Having knowledge of the Slater decomposed form allows for a quicker determination of various quantities related to entanglement and correlations. Furthermore, in some cases, the Slater decomposed form can be expressed in a simple analytical manner, enabling the derivation of analytical results for these quantities.

Understanding many low-energy phenomena in atomic nuclei greatly depends on rotational and isospin symmetry, as well as pairing correlations. The study of isovector (IV) pairing interactions has a long history, and the significance of the isoscalar (IS) pairing interaction between protons and neutrons is explored in references \cite{fra14, sag16, neg18}. The IV pairing interaction is essential for effective nuclear interactions, and its importance cannot be overstated.

Given the diverse range of effective interactions in nuclear physics, understanding their impact on mode entanglement is crucial for advancing the field.
In our previous work \cite{Kruppa21}, we utilized the USD \cite{Brown88} 
and USDB \cite{Brown06} interactions, where the two-body matrix elements were fitted to experimental data. In this research, we will investigate the entanglement properties of two-nucleon systems under the influence of either IV or IS pairing interactions.

This study provides a detailed examination of one-mode entropies and a comprehensive analysis of mutual informations for both IV and IS interactions, within the frameworks of both $jj$ coupling and $LS$ coupling. Our research highlights the significant influence of nuclear shell structure on one-mode entropies and mutual informations. We prove intriguing properties of the IV and IS interactions; specifically, in certain  ground state wave functions, each mode exhibits the same one-mode entropy, and the total correlation reaches its maximum value in the case of degenerate single-particle energies.

This work is organized as follows. In section \ref{section2}, the applied formalism of mode entanglement is reviewed. 
Section \ref{gen} discusses the general case of an arbitrary number of nucleons. It also reveals the consequences of symmetries in the wave function.
Section \ref{section4} reviews the Slater decomposed form of two-nucleon wave functions and its connection to one-mode entropies and mutual informations. In section \ref{section5}, single configurations of angular momentum and isospin coupled two-nucleon states 
are analysed, and then in section \ref{ciformsec} discusses the mixing of configurations. In section \ref{section7}, numerical results related to the sd shell are presented in the presence of pairing interactions. 
Finally, section \ref{section8}  summarizes our results.

\section{Mode entanglement and two-mode correlation}
\label{section2}

This section summarizes some concepts of mode entanglement and two-mode correlation,
which are important from the viewpoint of the present work. For the sake of simplicity, we define  these concepts here
only for the case of pure states with fixed number of nucleons. 

The primary objects of second quantized formalism are the creation and annihilation operators 
that refer to elements of an orthonormal basis of the single-particle (sp) Hilbert space. 
In this work, we consider finite dimensional sp Hilbert space with dimension $d$.
The elements of an orthonormal basis that spans this space are referred as modes 
and they are indexed from 1 to $d$. 
In the present study we discuss the mode entanglement in isospin formalism,
 so $d$ is the sum of the proton modes and the neutron modes.

The creation and the annihilation operator that corresponds to the mode with index $i$
 is denoted by $c^{\dagger}_i$ and $c_i$, respectively, and the vacuum of these operators is denoted by  
$\vert 0\rangle$.
The algebra of the operators is determined by  the canonical anticommutation relations
\begin{eqnarray}
c_ic_j^\dagger+c_j^\dagger c_i=\delta_{i,j},\nonumber\\
c_ic_j+c_jc_i=c^\dagger_ic_j^\dagger+c^\dagger_jc_i^\dagger=0.
\end{eqnarray}

The definition of mode entanglement is based on a bipartition of the modes \cite{Benatti20}. 
Here we consider such bipartitions, whose one part contains only one mode or two modes, and the other modes belong 
to the other part of the bipartition. So for these bipartitions 
the characterization of the mode entanglement is based on the one-mode reduced density matrix (1M-RDM) 
with respect to the given mode, 
or on the two-mode reduced density matrix (2M-RDM)  with respect to the given pair of modes, respectively.

These matrices can be defined easiest by expected values of operators in second quantized formalism.
For a wave function $\vert \psi\rangle$ that possesses fixed nucleon number, 
the 1M-RDM (with respect to the mode $i$) in the basis $\{c_i^\dagger \vert 0\rangle ,\vert 0\rangle\}$  takes the form 
\begin{equation}\label{onemden}
\rho^{\rm 1M-RDM}_i=\left(\begin{array}{cc}
\langle c_i^\dagger c_i\rangle&0\\
0& 1-\langle c_i^\dagger c_i\rangle
\end{array}\right),
\end{equation} 
 and 
the 2M-RDM (with respect to the pair composed by the modes $i$ and $j$) in the basis $\{c_i^\dagger 
c_j^\dagger|0\rangle ,c_i^\dagger |0\rangle , c_j^\dagger|0\rangle, |0\rangle\}$ has the form \cite{tul18,sav,ser17}
\beq
\rho^{\rm{2M-RDM}}_{(i,j)}=
\left(
\begin{array}{cccc}
\langle c_i^\dagger c_i c_j^\dagger c_j \rangle & 0 &0 & 0 \\
0& \langle c_i^\dagger c_i  c_j c_j^\dagger \rangle &\langle  c_j^\dagger c_i \rangle & 0 \\
0&  \langle c_i^\dagger  c_j \rangle &\langle  c_i c_i^\dagger c_j^\dagger c_j \rangle & 0 \\
0& 0 &0 & \langle c_i c_i^\dagger   c_j c_j^\dagger \rangle \\
\end{array}
\right).
\eeq
Here the notation $\langle O \rangle\equiv\langle \psi |O|\psi\rangle$ was introduced 
for the expected value of the operator $O$.
 
A further important concept is the one-particle reduced density matrix (1P-RDM), 
whose elements are defined here by the formula
\begin{equation}
\rho^{\rm 1P-RDM}_{i,j}=\langle\psi\vert c_j^\dagger c_i\vert\psi\rangle,
\end{equation}
which applies the normalization condition that the trace of 1P-RDM equals to the particle number.

The entropy of a given reduced density matrix provides a quantitative measure for the entanglement 
between the subsystems. 
In this work we use
the von Neumann entropy
\bea
S(\rho)=\rm{Tr~}{\rm H}(\rho), 
\eea
where the function H is defined as  
\bea
{\rm H} (x)\equiv \Bigg \{ 
\begin{array}{cc}
-x \log_2 (x) &{\rm ~~for~~} x>0,  \\ 
0 &{\rm ~~for~~} x=0.
\end{array}
\eea

The entropies of a 1M-RDM or a 2M-RDM are referred as one-mode and two-mode entropies, respectively.
The one-mode entropy with respect to mode $i$ is denoted by $S^{\rm 1M}_i$ and it can be expressed  as  
\bea\label{sidef}
S^{\rm 1M}_i\equiv S(\rho^{\rm 1M-RDM}_i)={\rm h}(\langle c_i^\dagger c_i\rangle)
={\rm h}\left(\rho^{\rm 1P-RDM}_{i,i}\right),
\eea
where the function $\rm{h}$ is defined as 
\bea\label{kish}
{\rm h} (x)\equiv {\rm H} (x)+{\rm H} (1-x).
\eea
The two-mode entropy with respect to mode $i$ and $j$
is denoted by $S^{\rm 2M}_{i,j}$ and it possesses the symmetry property $S^{\rm 2M}_{i,j}=S^{\rm 2M}_{j,i}$.

From the trace of the mode reduced density matrices we get a universal upper bound for the entropies. Namely, these upper limits of the one-mode and two-mode entropies are one and two, respectively. The one-mode entropy is determined entirely by the mode's occupation number. The upper limit of the one-mode entropy can be reached only if the occupation number of the mode is  $\frac{1}{2}$.

One-mode entropies characterize the wave function only from the viewpoint of a given mode. 
It is worth to determine one-mode entropies 
with respect to every mode of the given sp basis and study the sum of one-mode 
entropies  \cite{leg04}. 
Following \cite {ste18} we apply the term total correlation for the sum
\bea
I_{\rm tot}\equiv\sum_i S^{\rm 1M}_i.
\eea
This quantity depends on the choice of the sp basis. 

For a basis-independent characterization, 
one-body entanglement entropy \cite{gig15,gig16} is introduced by the definition
\beq
S^{\rm{1B}}={\rm min }~ I_{\rm tot},
\eeq
where the minimum is over all orthonormal bases of the sp Hilbert space.

The one-body entanglement entropy can be calculated in the form  \cite{gig15} 
\beq\label{onebent}
S^{\rm{1B}}=\sum_{i=1}^{d} {\rm h} (n_i),
\eeq 
where $n_i$ are the eigenvalues of the 1P-RDM i.e. the occupation numbers of the natural orbitals.

Mutual information can be introduced for the characterization of correlations between two modes.
Mutual informations between two different modes indexed by $i$ and $j$ is defined as  \cite{sza15}
\beq
I_{i,j}=S^{\rm 1M}_{i}+S^{\rm 1M}_{j}-S^{\rm 2M}_{i,j}.
\eeq
Since $S^{\rm 2M}_{i,j}=S^{\rm 2M}_{j,i}$, 
the value of the mutual information is also independent from the order of its indices. 

\section{Consequences of rotational symmetries in space and in isospin space}
\label{gen}

Before the discussion of two-nucleon systems 
it is merit to pay attention to some general consequences  of symmetries 
that are valid for systems with an arbitrary number of particles. 
In this section we review some consequences of symmetries that
correspond to rotations in space or in isospin space. 
For the sake of simplicity here we consider wave functions that are characterized by
given quantum numbers concerning the total angular momentum, the total isospin and
their projections. Here these quantum numbers are denoted by $J$, $T$, $J_z$ and $ T_z$, respectively.

In this section the symmetry properties are investigated in an sp basis of the $jj$ coupling scheme and
the creation operators of the basis elements are denoted as $c^\dagger_{\rho, j,j_z,\frac{1}{2},t_z}$. 
A creation operator
$c^\dagger_{\rho,j,j_z,\frac{1}{2},t_z}$ creates a nucleon with quantum numbers $j$ for
 angular momentum, $j_z$ for angular momentum projection, $\frac{1}{2}$ for isospin and $t_z$ for isospin projection.
For the unique identification of the orbital we introduce the index $\rho$, 
so that $c^\dagger_{\rho,j,j_z,\frac{1}{2},t_z}$ is the $j_z$-th component of a rank $j$ spherical tensor operator 
and the $t_z$-th component of a rank $\frac{1}{2}$ isospin tensor. 
Henceforward the notation  $c_{\rho,j,\frac{1}{2}}^\dagger$ is applied for the related double-tensor.

Here we discuss three symmetry properties of the one-mode entropies.
Firstly, it is shown that when the projection of the total angular momentum is zero, the modes that differ 
only in the sign of $j_z$ have the same one-mode entropy.
Secondly, we prove that when the projection of the total isospin is zero, 
the one-mode entropies do not depend on the  $t_z$ quantum number of the modes. 
Finally, it is shown that for angular momentum scalar wave functions 
one-mode entropies are independent from the  $j_z$ quantum number of the modes. 

We introduce the modified annihilation operator
\bea
\tilde c_{\rho,j,j_z,\frac{1}{2},t_z} \equiv (-1)^{j+j_z+\frac{1}{2}+t_z}c_{\rho,j,-j_z,\frac{1}{2},-t_z},
\eea
which is also a $j_z$-th component of a rank $j$ spherical tensor operator 
and a $t_z$-th component of a rank $\frac{1}{2}$ isospin tensor. 
In this work the corresponding double-tensor is denoted 
as $\tilde c_{\rho,j,\frac{1}{2}}$ and for the coupling of tensors $c_{\rho_1,j_1,\frac{1}{2}}^\dagger$ and  
$\tilde c_{\rho_2,j_2,\frac{1}{2}}$ the notation
$\left [c_{\rho_1,j_1,\frac{1}{2}}^\dagger \otimes \tilde c_{\rho_2,j_2,\frac{1}{2}}\right]^{\kappa,\tau}_{\kappa_z,\tau_z}
=\underset{j_z,j'_z}{\sum}\underset{t_z,t'_z}{\sum}C^{\kappa,\kappa_z}_{j_1,j_z,j_2,j'_z}
C^{\tau,\tau_z}_{\frac{1}{2},t_z,\frac{1}{2},t'_z}
c_{\rho_1,j_1,j_z,\frac{1}{2},t_z}^\dagger \tilde c_{\rho_2,j_2,j'_z,\frac{1}{2},t'_z}$
is applied, where the Clebsch-Gordan coefficients are denoted in the usual way.

Henceforward  $\Psi^{J,T}_{J_z,T_z}$ signs an arbitrary wave function that is characterized by
the earlier mentioned four quantum numbers. To determine the one-mode entropies of $\Psi^{J,T}_{J_z,T_z}$
we have to calculate the matrix elements 
$\langle\Psi^{J,T}_{J_z,T_z}\vert c_{\rho,j,j_z, \frac{1}{2},t_z}^\dagger  c_{\rho,j,j_z, \frac{1}{2},t_z}
\vert\Psi^{J,T}_{J_z,T_z}\rangle$.
 
The completeness property of Clebsch-Gordan coefficients implies
\bea
c_{\rho,j,j_z,\frac{1}{2},t_z}^\dagger \tilde c_{\rho,j,-j_z,\frac{1}{2},-t_z}=
\underset{\kappa,\tau} {\sum}
C^{\kappa,0}_{j,j_z,j,-j_z}
C^{\tau,0}_{\frac{1}{2},t_z,\frac{1}{2},-t_z} 
\left[c_{\rho,j,\frac{1}{2}}^\dagger\otimes \tilde c_{\rho,j,\frac{1}{2}}\right]^{\kappa,\tau}_{0,0}, 
\eea
so the application of Wigner-Eckart theorem to the double-tensors leads to
\bea\label{onemodecalc} 
&&\langle\Psi^{J,T}_{J_z,T_z}
\vert c_{\rho,j,j_z, \frac{1}{2},t_z}^\dagger  c_{\rho,j,j_z, \frac{1}{2},t_z}\vert\Psi^{J,T}_{J_z,T_z}\rangle
\nonumber \\
&&=\frac{(-1)^{j-j_z+\frac{1}{2}-t_z}}{\sqrt{(2J+1)(2T+1)}} 
\sum_{\kappa,\tau}\Bigg(
C^{\kappa,0}_{ j,j_z,j,-j_z}
C^{\tau,0}_{\frac{1}{2},t_z,\frac{1}{2},-t_z}
C^{J,J_z}_{J,J_z,\kappa,0}C^{T,T_z}_{T,T_z,\tau,0} \nonumber \\
&&\times \langle\Psi^{J,T}\vert \vert\vert
 \left [c_{\rho,j,\frac{1}{2}}^\dagger\otimes \tilde c_{\rho,j,\frac{1}{2}}\right]^{\kappa,\tau}
\vert \vert\vert\Psi^{J,T}\rangle \Bigg),
\eea
where we introduced the double reduced matrix elements.
All of the mentioned symmetry properties of one-mode entropies are the consequences of this formula.

Firstly, we consider the case when the wave function possesses axial symmetry around the $z$-axis, i.e. $J_z=0$. 
In this case $J$ is integer, but there is no any further restriction to the values of the other three quantum numbers.
The coefficient $C^{J,J_z=0}_{J,J_z=0,\kappa,0}$ can be nonzero only if $\kappa$ is even, 
 so the summation over $\kappa$  in (\ref{onemodecalc}) can be restricted to these cases. 
If  $\kappa$  is even, 
$C^{\kappa,0}_{ j,j_z,j,-j_z}=-C^{\kappa,0}_{j,-j_z,j,j_z}$.
In (\ref{onemodecalc}) only the factor $(-1)^{j-j_z+\frac{1}{2}-t_z}$ and the coefficient 
$C^{\kappa,0}_{ j,j_z,j,-j_z}$ depend on $j_z$.  
This means  that for $J_z=0$  the $j_z \to -j_z$ substitution in 
(\ref{onemodecalc})  changes only the 
sign of the terms $(-1)^{j-j_z+\frac{1}{2}-t_z}$ and $C^{\kappa,0}_{ j,j_z,j,-j_z}$,
so this substitution does not change the value of $\langle\Psi^{J,T}_{J_z=0,T_z}
\vert c_{ \rho,j,j_z, \frac{1}{2},t_z}^\dagger  c_{ \rho,j,j_z, \frac{1}{2},t_z}\vert\Psi^{J,T}_{J_z=0,T_z}\rangle$ . 
This means that in the $J_z=0$ case  the one-mode entropies do not depend 
on the sign of the $j_z$ quantum number of the mode. 

Secondly we discuss the analog of the previous symmetry in isospin space, that is we
consider the $T_z=0$ case without any further restriction for the values of $J$, $J_z$ and $T$.
In this case the  proton and neutron numbers are the same and these systems possess great physical importance. 
Since $C^{T,0}_{T,0,\tau,0}$ can be nonzero only when $2T+\tau$ is even, 
in our case  (since $T_z=0$ implies that $T$ is integer) $C^{T,0}_{T,0,\tau,0}$ 
can be nonzero only when $\tau$ is even.
Since $\tau$ can take only two values (zero and one), only the $\tau=0$ term can give contribution to the summation 
 over $\tau$  in (\ref{onemodecalc}).
With the use of the identities  
$C^{\tau=0,0}_{ \frac{1}{2},t_z,\frac{1}{2},-t_z}=\frac{(-1)^{\frac{1}{2}-t_z}}{\sqrt{2}}$ 
and $C^{T,T_z}_{T,T_z,\tau=0,0}=1$
we can obtain the formula
\bea\label{onemodecalc1}
&&\langle\Psi^{J,T}_{J_z,T_z=0}
\vert c_{ \rho,j,j_z, \frac{1}{2},t_z}^\dagger  c_{ \rho,j,j_z, \frac{1}{2},t_z}\vert\Psi^{J,T}_{J_z,T_z=0}\rangle
\nonumber \\
&&=\frac{(-1)^{j-j_z}}{\sqrt{2(2J+1)(2T+1)}}\sum_{\kappa}\Bigg(
C^{\kappa,0}_{ j,j_z,j,-j_z}C^{J,J_z}_{J,J_z,\kappa,0} \nonumber \\
&&\times \langle\Psi^{J,T}\vert \vert\vert
 \left [c_{ \rho,j,\frac{1}{2}}^\dagger\otimes \tilde c_{ \rho,j,\frac{1}{2}}\right]^{\kappa,\tau=0}\vert 
\vert\vert\Psi^{J,T}\rangle \Bigg).
\eea
This formula shows that the one-mode entropies do not depend on the $t_z$ quantum number of the mode,
if the number of protons and the number of neutrons are the same. 

Finally, we discuss the case of spherical scalar wave function, when $J=J_z=0$. 
Since $C^{0,0}_{0,0,\kappa,0}=\delta_{\kappa,0}$, 
only the $\kappa=0$ term remains in  the summation of (\ref{onemodecalc}). 
If we substitute  $C^{\kappa=0,0}_{ j,j_z,j,-j_z}$  with its value $(-1)^{j-j_z}/\sqrt{2j+1}$ 
we can derive the formula
\bea\label{onemodecalc2}
&&\langle\Psi^{J=0,T}_{J_z=0,T_z}
\vert c_{ \rho,j,j_z, \frac{1}{2},t_z}^\dagger  c_{ \rho,j,j_z, \frac{1}{2},t_z}\vert\Psi^{J=0,T}_{J_z=0,T_z} \rangle
\nonumber \\
&&=\frac{(-1)^{\frac{1}{2}-t_z}}{ \sqrt{(2j+1)(2T+1)}}\sum_\tau\Bigg(
C^{\tau,0}_{ \frac{1}{2},t_z,\frac{1}{2},-t_z} C^{T,T_z}_{T,T_z,\tau,0}
\nonumber\\
&&\times \langle\Psi^{J=0,T}\vert \vert\vert
 \left [c_{ \rho,j,\frac{1}{2}}^\dagger\otimes \tilde c_{ \rho,j,\frac{1}{2}}\right]^{\kappa=0,\tau}
\vert \vert\vert\Psi^{J=0,T}\rangle \Bigg).
\eea
This expression shows that for states with zero total angular momentum the one-mode entropies do
not depend on the $j_z$ quantum number of the given mode. 
We mentioned that this symmetry property is also noted in a previous work \cite{Kruppa22}, 
but there the proof was not detailed.

Obviously this symmetry property also has an analog in isospin space, but its analog 
(the $T=0$ case) does not provide consequences  that go beyond the earlier discussed $T_z=0$ case.
 
\section{Two-nucleon wave functions in Slater decomposed form}
\label{section4}

An arbitrary two-nucleon pure state can be represented as 
\beq
 \vert\psi \rangle
=\sum_{i,j=1}^d w_{i,j} c_{i}^\dagger c_{j}^\dagger|0\rangle,
\eeq
where the coefficients $w_{i,j}$ are the elements of a skew symmetric $d \times d$ matrix $w$.
If we apply the real normal form of antisymmetric matrices \cite{fiz2} 
we can write $\vert \psi \rangle$ in the so-called Slater decomposed form \cite{sch01}
\beq\label{sf}
 \vert\psi \rangle
=2\sum_{s=1}^n \lambda_{s} a_{2s-1}^\dagger a_{2s}^\dagger|0\rangle,
\eeq
where $a^\dagger_i$ denotes the creation operator belongs to the $i$-th element of a new sp basis, 
which we call Slater basis. 
The coefficients $\lambda_s$ are positive real numbers, 
and $n$, the upper limit of the summation in (\ref{sf}), is the Slater rank of 
the state \cite{eck02}. Applying the term introduced in \cite{Kruppa21},
the members of a pair of modes $2s-1$ and $2s$ in the Slater decomposed form are said to be associated.
The Slater basis is a natural basis,  since in this basis 
$\rho^{\rm{1P-RDM}}$ is diagonal \cite{Kruppa21}. 
The nonzero elements of $\rho^{\rm{1P-RDM}}$ in the main diagonal are $4\lambda^2_{s}$.
The eigenvalues of $\rho^{\rm{1P-RDM}}$ in the Slater basis are at least two-fold degenerate
since a pair of modes belongs to every coefficient $\lambda_s$.
Zero eigenvalues occur only if $2n<d$. 

It is practical to express the quantities that characterize the wave function by the coefficients $\lambda_s$.
In the Slater basis,  members of an associated mode pair have the same one-mode entropies  
\bea\label{onemodeval}
S^{\rm 1M}_{2s-1}=S^{\rm 1M}_{2s}={\rm h} (4\lambda^2_s),
\eea
and there are $d-2n$ modes with zero one-mode entropy, since they do not occur in the wave function. 
The one-mode entropies are determined by the coefficients of the Slater decomposition. 
The closer the $2\lambda$ value of the mode to  $\frac{1}{\sqrt 2}$ the larger the corresponding one-mode entropy.

The Slater decomposed form  (\ref{sf}) groups into pairs the $2n$ modes of the Slater basis 
that present in the wave function. 
The $s(i)$ notation in the following formulas means that  
the index $i$ of a mode determines the pair that contains this mode, 
so it determines the index of $\lambda$ corresponding to this pair.
A Slater basis is natural basis, the nonzero  occupation numbers are $n_i=4\lambda^2_{s(i)}$ so according to (\ref{onebent}) the one-body entanglement entropy is
\bea
S^{\rm{1B}}=\sum_{s=1}^{n}2  {\rm h}(4\lambda^2_s). 
\eea
The two-mode entropies also have simple expressions in the Slater basis:
\bea\label{twome}
S^{\rm 2M}_{i,j}=\left\{
\begin{array}{l}
{\rm h}(4\lambda_{s(i)}^2)$ if $i$ and $j$ are associated$, \\
{\rm H}(4\lambda_{s(i)}^2)+{\rm H}(4\lambda_{s(j)}^2)+{\rm H}(1-4\lambda_{s(i)}^2-4\lambda_{s(j)}^2) 
$ if both modes$ \\
$are present in the state, but they are not associated$, \\
{\rm h}(4\lambda_{s(k)}^2) $ if only mode (denoted as $k$) of the pair $(i,j)$ is$ \\
$present in the state$, \\
0$ if none of the modes are present in the state$.
\end{array}
\right. 
\eea
In the case when the modes $i$ and $j$ form a pair of associated modes i.e. $s(i)=s(j)$, 
the one-mode entropies and two-mode entropies coincide
\bea\label{assval}
S^{\rm 2M}_{i,j}=S^{\rm 1M}_{i}=S^{\rm 1M}_{j}={\rm h}(4\lambda_{s(i)=s(j)}^2).
\eea

If the modes $i$ and $j$ are modes of a Slater basis, 
the largest value of $S^{\rm 2M}_{i,j}$ that can be reached is 
$\log_2 (3)\approx 1.58$. This limit can be reached if the modes are not associated and
 the occupation number of both mode is $\frac{1}{3}$.
For a nonassociated mode pair $(i,j)$, the expected value of the product of the occupation number operator for mode $i$ and the occupation number operator for mode $j$ is always zero. Consequently, when considering nonassociated mode pairs from the Slater basis, the two-mode reduced density matrix can have at most three nonzero diagonal elements. The mentioned upper limit can be reached when all the three nonzero values in the diagonal of the density matrix are equal.   

The mutual information in the Slater basis can be turned into the form
\bea\label{mifval}
I_{i,j}=\left\{
\begin{array}{l}
{\rm h}(4\lambda_{s(i)}^2)={\rm h}(4\lambda_{s(j)}^2))$ if $i$ and $j$ are associated$, \\
{\rm H} (1-4\lambda^2_{s(i)})+{\rm H} (1-4\lambda^2_{s(j)})-{\rm H}(1-4\lambda_{s(i)}^2-4\lambda_{s(j)}^2) 
$ if both$ \\
$modes are present in the state, but they are not associated$, \\
0 $ if one of the modes is not present in the state$.
\end{array}
\right. 
\eea
From (\ref{assval}) and (\ref{mifval}) it follows, 
that the mutual information of an associated mode pair is the same as the one-mode entropies 
of the constituting modes of the pair. 
In \ref{app3} we prove that in Slater basis the mutual information of the associated modes 
is the largest among the mutual information values in the sense that
$I_{i, i_{\rm ass}}=\underset{j}{\max}\ I_{i,j}$, 
where the modes $i$ and  $i_{\rm ass}$ constitute an associated pair.

\section{States with single configuration}
\label{section5}

In this section we work in sp basis of the $jj$ coupling scheme.
We give the two-particle states in isospin formalism, 
so protons and neutrons are considered as different states of the nucleon.
Henceforward a mode $i$ are defined by the quantum numbers $ \rho$,$j$,$j_z$ and $t_z$.
An unnormalized two-nucleon wave function, which is  determined by the usual angular momentum and isospin couplings of two orbitals 
is $\left [ c^\dagger_{ \rho_1,j_1,\frac{1}{2}}\otimes c^\dagger_{ \rho_2,j_2,\frac{1}{2}}\right ]^{J,T}_{J_z,T_z}
\vert 0\rangle$. Its normalized form is what we refer to as a single configuration:
\begin{eqnarray}
&&\psi^{{J,J_z,T,T_z}}_{ \rho_1,j_1, \rho_2,j_2}=N^{J,J_z,T,T_z}_{ \rho_1,j_1, \rho_2,j_2} \times \nonumber \\
&&\sum_{{j_z}\in{\cal S}_{j_1,j_2,J_z}}\sum_{t_z}
C^{J,J_z}_{j_1,j_z,j_2,\overline{j_z}}C^{T,T_z}_{\frac{1}{2},t_z,\frac{1}{2},T_z-t_z} 
c_{ \rho_1,j_1,j_z,t_z}^\dagger c_{ \rho_2,j_2,\overline{j_z},T_z-t_z}^\dagger|0\rangle. 
\end{eqnarray}
Here we introduced the notation $\overline{j_z}\equiv J_z-j_z$, the set ${\cal S}_{j_1,j_2,J_z}$, which  is defined as ${\cal S}_{j_1,j_2,J_z}\equiv\{j_z|2j_z 
{\rm\ is\ odd}, 
-j_1\le j_z \le j_1 {\rm\ and } -j_2\le \overline{j_z} \le j_2\}$ and the normalization 
condition is 
\begin{equation}
N^{J,J_z,T,T_z}_{ \rho_1,j_1, \rho_2,j_2}
=\frac{\sqrt{1+\delta_{ \rho_1, \rho_2}\delta_{j_1,j_2}}}{1+\delta_{ \rho_1, \rho_2}
\delta_{j_1,j_2}}.
\end{equation}

Henceforward the term same subshells is applied for the cases, when
 $ \rho_1= \rho_2$ and $j_1=j_2$,
and the term different subshells is applied for the other cases, when $ \rho_1\neq \rho_2$ or $j_1\neq j_2$.
The Slater decomposed form of the coupled wave functions can be read out easily
by the rearrangements of the modes and the terms of the wave functions. 
For example, for same subshells the isoscalar 
wave function with odd $J$ is  
\bea\label{isoscaformsame}
\psi^{J,J_z,0,0}_{ \rho,j, \rho,j}=
-\underset{j_z \in {\cal S}_{j,j,J_z}}{\sum}
C^{J,J_z}_{j,j_z,j,\overline {j_z}}
c_{ \rho,j,j_z,-\frac{1}{2}}^\dagger c_{ \rho,j,\overline {j_z},\frac{1}{2}}^\dagger|0\rangle
\eea
and for different subshells with arbitrary $J$ is 
\bea\label{isoscaformdiff}
\psi^{J,J_z,0,0}_{ \rho_1,j_1, \rho_2,j_2}&=& \frac{1}{\sqrt{2}}
\underset{{j_z}\in{\cal S}_{j_1,j_2,J_z}}{\sum}
C^{J,J_z}_{j_1,j_z,j_2,\overline{j_z}}\bigg[
c_{ \rho_2,j_2,\overline{j_z},\frac{1}{2}}^\dagger c_{ \rho_1,j_1,j_z,-\frac{1}{2}}^\dagger |0\rangle \nonumber\\
&&+ c_{ \rho_1,j_1,j_z,\frac{1}{2}}^\dagger c_{ \rho_2,j_2,\overline{j_z},-\frac{1}{2}}^\dagger|0\rangle\bigg].
\eea
These states are not in Slater decomposed form, but it's simple to identify the associated pairs and their coefficients. The $\lambda$ coefficients are proportional to the absolute values of the nonzero Clebsch-Gordan coefficients.
The similar formulas for the isovector cases are given in \ref{app1} and
the results are summarized in table \ref{table1}.

\begin{table}[h]
\caption{\label{table1}Characteristics of the angular momentum and isospin coupled single configurations. 
The Slater rank, occupation number $4\lambda^{2}(j_z)$ and the associated modes are given. 
In the case of associated modes, we assume that none of the Clebsch-Gordan coefficients in the wave function  are exceptional ones.
In the case of identical shells due to the antisymmetry requirement the restriction  
 $(-1)^{J+T}=-1$  is imposed. For further explanation see the text. }
\begin{indented}
\item[]\begin{tabular}{cccccc}
\br
T&$T_z$  & subshells   & Slater rank &  $4\lambda^{2}(j_z)$ & associated pairs\\ 
\mr
1&$\pm 1$  & same   &$\frac{{ R}_{j,j,J,J_z}}{2} $ 
& $2\left(C^{J,J_z}_{j,j_z,j,\overline {j_z}}\right)^2$ &   $\left( \rho,j,j_z,\pm \frac{1}{2}; \rho,j,\overline {j_z}, \pm 
\frac{1}{2} \right)$ \\
\hline
1&$\pm 1$ & different    & ${R}_{j_1,j_2,J,J_z}$ 
& $
\left(C^{J,J_z}_{j_1,j_z,j_2,\overline{j_z}}\right)^2 $ 
& $\left( \rho_1,j_1,j_z,\pm \frac{1}{2}; \rho_2,j_2,\overline {j_z}, 
\pm \frac{1}{2} \right)$ \\
\hline
1&0 & same   & ${R}_{j,j,J,J_z}$   
& $\left(C^{J,J_z}_{j,j_z,j,\overline {j_z}}\right)^2 $ 
& $\left( \rho,j,j_z, -\frac{1}{2}; \rho,j,\overline {j_z}, +\frac{1}{2} 
\right)$\  \\
\hline
\multirow{2}*{1}&\multirow{2}*{0} & \multirow{2}*{different}  &\multirow{2}*{ $2{R}_{j_1,j_2,J,J_z} $}
&\multirow{2}*{ $\frac{1}{2}\left(C^{J,J_z}_{j_1,{j_z},j_2,\overline{j_z}}\right)^2 $ }& $\left( \rho_1,j_1,j_z,-\frac{1}{2}; \rho_2,j_2,
\overline {j_z}, +\frac{1}{2} \right)$ \\
& & & & &  $\left( \rho_1,j_1,j_z,+\frac{1}{2}; \rho_2,j_2,\overline {j_z}, -\frac{1}{2} \right)$ \\
\hline
0&0 & same  
& ${R}_{j,j,J,J_z}  $ 
 & $\left(C^{J,J_z}_{j,j_z,j,\overline {j_z}}\right)^2$ 
&  $\left( \rho,j,j_z, -\frac{1}{2}; \rho,j,\overline {j_z}, +\frac{1}{2} \right)$\\
\hline
\multirow{2}*{0}&\multirow{2}*{0} &\multirow{2}*{different}
&\multirow{2}*{$2{R}_{j_1,j_2,J,J_z} $} 
&\multirow{2}*{ $\frac{1}{2}\left(C^{J,J_z}_{j_1,j_z,j_2,\overline{j_z}}\right)^2 $}  & $\left( \rho_1,j_1,j_z,-\frac{1}{2}; \rho_2,j_2,
\overline {j_z}, +\frac{1}{2} \right)$ \\
& & & & &   $\left( \rho_1,j_1,j_z,+\frac{1}{2}; \rho_2,j_2,\overline {j_z}, -\frac{1}{2} \right)$ \\
\br
\end{tabular}
\end{indented}
\end{table}

In the determination of the Slater rank in table \ref{table1} we considered the following.  
The Clebsch-Gordan coefficients have a peculiarity that there are so-called exceptional cases,
where the quantum numbers satisfy all  familiar conditions, 
nevertheless, the value of the Clebsch-Gordan coefficient is zero \cite{ray78, hei09}. 
The set of $j_z$ quantum numbers of this kind of Clebsch-Gordan coefficients $C^{J,J_z}_{j_1,j_z,j_2,\overline{j_z}}$ 
is denoted by ${\cal S}^{{\rm exceptional}}_{j_1,j_2,J,J_z}$. 
We introduce the notation 
${R}_{j_1,j_2,J,J_z}=|{\cal S}_{j_1,j_2,J_z}|-|{\cal S}^{{\rm exceptional}}_{j_1,j_2,J,J_z}|$ and 
we sign the cardinality of the set $\cal S$ by  $|\cal S|$.

From table \ref{table1} it can be observed that the switch from the $T_z=\pm 1$ cases to the $T_z=0$ cases
or the switch from same to different subshells doubles the Slater rank .
It is worth to note that the case of $J=0$ leads to significant simplification, since 
$C^{0,0}_{j_1,m_1,j_2,-m_2}
=\delta_{j_1,j_2}\delta_{m_1,m_2}(-1)^{j_1-m_1}/\sqrt{2j_1+1}$.
For this case the obtained analytical results are given in \ref{app1}.

In the case of single configurations, we can study the optimal choice for maximizing the one-mode entropies. From table 1 and table A1 it follows that the one-mode entropy reaches the theoretical upper bound in several cases. For example, in case of identical subshells the one-mode entropy of all modes of the single configuration  with quantum numbers $J=0,\ T=1,\ T_z=0,\ j=\frac{1}{2}$  equal one.

\section{\label{ciformsec}States with configuration mixing}

In this section we discuss more general states than single configurations, when the full 
wave function is a superposition of single configurations. Henceforward we 
will refer to these mixed configurations as configuration interaction (CI) wave functions. 

In special cases the Slater decomposed form of CI wave function 
can be determined without numerical calculations. 
Such case is realized if the following two conditions are satisfied:
the CI two-nucleon state is spherical scalar and 
a given sp angular momentum $j$ uniquely determines the quantum number $ \rho$. 
We will discuss such cases, so in this section 
the $\rho$ index of creation and annihilation operators are abandoned. 

Henceforward, for brevity the  $\psi_{j,T_z}$  notation is introduced for single configurations 
with  quantum numbers $J=J_z=0$, $T=1$, $j_1=j_2=j$ and $T_z$, that is 
\bea
\psi_{j,T_z=0}&=& 
\sum_{j_z \in {\cal S}_{j,j,0}}C^{0,0}_{j,j_z,j,\overline{j_z}}
c_{j,j_z,-\frac{1}{2}}^\dagger c_{j,\overline{j_z},\frac{1}{2}}^\dagger|0\rangle, \nonumber \\
 \psi_{j,T_z=\pm 1}&=& 
\sqrt{2}
\underset{j_z \in {\cal S}_{j,j,0},j_z<0}
{\sum}
C^{0,0}_{j,j_z,j,\overline{j_z}}
c_{j,j_z,\pm\frac{1}{2}}^\dagger c_{j,\overline{j_z},\pm\frac{1}{2}}^\dagger|0\rangle.
\eea
In the investigated case we can write a spherical scalar wave function in the form
\bea\label{ciform}
\Psi_{T_z}=\sum_{j}A_{j,T_z}\psi_{j,T_z}.
\eea
Without loss of generality, we assume that coefficients $A_{j,T_z}$ are real and nonzero.
To analyse states of the form (\ref{ciform}) from the viewpoint of entanglement,
it is worth to start with the determination of the coefficients that occur in their Slater decomposed form.
Due to the restrictions on the components of the superposition (\ref{ciform}) 
the $\lambda_s$ coefficients of the mixed configurations are 
\bea\label{note}
\lambda_s [\Psi_{T_z}]
=|A_{j(s),T_z}|\lambda[\psi_{j(s),T_z}]
=\left\{
\begin{array}{ll}
\frac{|A_{j(s),T_z}|}{2\sqrt{2j+1}} & $for $T_z=0,\\ 
\frac{|A_{j(s),T_z}|}{\sqrt{2}\sqrt{2j+1}} & $for $T_z=\pm 1,
\end{array}
\right.
\eea
where the $j(s)$ notation means that the index of $\lambda$ determines the value of the quantum number $j$.
The notation (\ref{note}) exploits that all $\lambda$ coefficients of a spherical scalar single configuration
is the same (see table \ref{table2}). 
From (\ref{onemodeval}) and (\ref{note}) it follows that 
the one-mode entropies of our CI wave function takes the form
\bea\label{cimode1}
S^{\rm 1M}_{k}[\Psi_{T_z}]={\rm h}\Big(n_k[\Psi_{T_z}]\Big)=
\left\{
\begin{array}{ll}
{\rm h} \left(\frac{A^2_{j(k),T_z}}{ 2j(k)+1}\right)   &$for $T_z=0, \\
{\rm h} \left(\frac{2A^2_{j(k),T_z}}{ 2j(k)+1}\right) &$for $T_z=\pm 1,
\end{array}
\right.
\eea
if the mode $k$ is an element of our Slater basis and present in the wave function (\ref{ciform}).
We can use (\ref{mifval}) and (\ref{note}) to write the formulas of table \ref{tablemi} for mutual informations.  
\begin{center}
\begin{table}[h]
\caption{\label{tablemi} Mutual informations in Slater basis for the investigated CI wave functions, 
when both members of the mode pair present in the wave function.}
\begin{indented}
\item[]\begin{tabular}{ccc}
\br
\multirow{2}*{$T_z$}&the mode&\multirow{2}*{$I_{k,k'}$}\\ 
        & pair $(k,k')$ & \\ 
\mr
$0$  & associated&${\rm h} \left(\frac{A^2_{j(k),T_z}}{ 2j(k)+1}\right)$ \\ 
$0$  &nonassociated   
& ${\rm H} \left(1-\frac{A^2_{j(k),T_z}}{ 2j(k)+1}\right)+{\rm H} \left(1-\frac{A^2_{j(k'),T_z}}{ 2j(k')+1}\right)
-{\rm H}\left(1-\frac{A^2_{j(k),T_z}}{ 2j(k)+1}-\frac{A^2_{j(k'),T_z}}{ 2j(k')+1}\right)$\\ 
$\pm 1$   &associated& ${\rm h} \left(\frac{2A^2_{j(k),T_z}}{ 2j(k)+1}\right)$\\ 
$\pm 1$  &nonassociated
& ${\rm H} \left(1-\frac{2A^2_{j(k),T_z}}{ 2j(k)+1}\right)+{\rm H} \left(1-\frac{2A^2_{j(k'),T_z}}{ 2j(k')+1}\right)
-{\rm H}\left(1-\frac{2A^2_{j(k),T_z}}{ 2j(k)+1}-\frac{2A^2_{j(k'),T_z}}{ 2j(k')+1}\right)$ \\
\br
\end{tabular}
\end{indented}
\end{table}
\end{center}

In this work we assume that if we list the modes of the sp model space, 
every ($j$, $j_z$) pair of quantum numbers occurs twice,
since both proton and both neutron mode belongs to a given ($j$, $j_z$) pair. 
In this case by (\ref{onebent}) and (\ref{cimode1}) we can obtain the one-body entanglement entropy 
of the investigated CI states as
\bea\label{onebodyentci}
S^{\rm{1B}}[\Psi_{T_z}]&=&
\left\{\begin{array}{ll}2\underset{j}{\sum}  (2j+1){\rm h}\left(\frac{A_{j,T_z}^2}{2j+1}\right) 
&{\ \rm for\ } T_z=0, \\
\underset{j}{\sum}(2j+1) 
 {\rm h}\left(\frac{2A^2_{j,T_z}}{2j+1}\right) 
&{\ \rm for\ } T_z=\pm 1.
\end{array}
\right.
\eea

If we fix the available modes we can search the wave function that gives the maximal one-body entanglement entropy.  
The one-body entanglement entropy is maximal if 
all occupation numbers of natural orbitals are identical and their values are 
${2}/{n_{\rm mod}}$,
where $n_{\rm mod}$ is the total number of the selected modes \cite{Kruppa21}. 

First we consider all modes, in this case $n_{\rm mod}=2\underset{j}{\sum}(2j+1)$,
and secondly we consider such modes, which have the same isospin projection, so 
in the second case $n_{\rm mod}=\underset{j}{\sum}(2j+1)$. 
In the first case $T_z=0$, in the second case $T_z=\pm 1$ for the wave function that maximize the entropy.
From the equation ${2}/{n_{\rm mod}}=4 (\lambda[\Psi_{T_z}])^2$ and (\ref{note})
we can determine the amplitudes of the CI wave function (\ref{ciform}),
which leads to maximal one-body entanglement entropy. 
 
From (\ref{onebodyentci}) we get that both for $T_z=0$ and $T_z=\pm 1$ the amplitudes 
that gives maximal one-body entanglement entropy are given by
\beq\label{maxent}
A_{j',T_z}^2=\frac{2j'+1}{\underset{j}{\sum} (2j+1)},
\eeq
so the maximal value is
\bea
S^{\rm{1B}}_{\rm max}&=&n_{\rm mod}{\rm h}\left(\frac{2}{n_{\rm mod}}\right).
\eea

Since this formula for $S^{\rm{1B}}_{\rm max}$ is a monotonically increasing function of $n_{\rm mod}$,
it leads to greater value in the $T_z=0$ case than in the $T_z=\pm 1$ case.
We can compare model spaces with and without isospin freedoms, concluding that the former leads to a larger upper limit for one-body entanglement entropy. 
This conclusion is similar, to some extent, to the result of \cite{Chen10},
where it is observed that consideration of new sp degrees of freedom can increment the entanglement length.  

\section{Isovector and isoscalar pairing interaction}
\label{section7}

In this section we study the two-nucleon problem in the case of the presence of IV and IS pairing interactions.
In the case of degenerate sp energies analytical result can be obtained, but in 
more general case numerical investigation may be needed, 
which is presented for the sd shell.
We apply different entanglement measures and study their behaviour in order to describe the correlation structure of
CI wave functions. 
Our numerical results are obtained by the application of 
the SNEG  library \cite{SNEG1,SNEG2} of Mathematica \cite{Mathematica}.

Following \cite{kot06} we define the pairing interaction in $LS$ coupling scheme, 
where the operator $a^\dagger_{n,l,m,\frac{1}{2},s_z,\frac{1}{2},t_z}$ 
creates a nucleon with orbital angular momentum $l$ and its projection $m$,
spin $\frac{1}{2}$ and its projection $s_z$, isospin   $\frac{1}{2}$ and its projection $t_z$, 
and  the time-reversed tensor operator is 
${\tilde a}_{n,l,m,\frac{1}{2},s_z,\frac{1}{2},t_z}
=(-1)^{l+m+1+t_z+s_z}a_{n, l,-m,\frac{1}{2}, -s_z,\frac{1}{2}, -t_z}$. 
We use the following isovector and isoscalar pair creation operators \cite{kot06} 
\beq\label{ivecop}
P^\dagger_{{\rm IV},T_z}=\sum_n \sum_l \sqrt{\frac{2l+1}{2}}\left[a^\dagger_{n,l,\frac{1}{2},\frac{1}{2}}
\otimes {a}^\dagger_{n,l,\frac{1}{2},\frac{1}{2}}\right]^{0,0,1}_{0,0,T_z}
\eeq
and
\beq\label{iscaop}
P^\dagger_{{\rm IS},S_z}=\sum_n \sum_l \sqrt{\frac{2l+1}{2}}\left[a^\dagger_{n,l,\frac{1}{2},\frac{1}{2}}
\otimes {a}^\dagger_{n,l,\frac{1}{2},\frac{1}{2}}\right]^{0,1,0}_{0,S_z,0}.
\eeq
The tensor coupling is denoted by $\otimes$ and the upper indices  are $L,\ S$ and $T$ (the quantum numbers of the coupled tensor)  
and the corresponding projections are signed as lower indices in (\ref{ivecop}) and (\ref{iscaop}).

The studied isovector and isocalar pairing interactions can be given in the standard way
\beq\label{isovecpot}
V^{\rm IV}=-G \sum_{T_z} P^\dagger_{{\rm IV},T_z} P_{{\rm IV},T_z}
\eeq
and
\beq\label{isoscapot}
V^{\rm IS}=-G\sum_{S_z} P^\dagger_{{\rm IS},S_z}P_{{\rm IS},S_z}.
\eeq

We consider the following Hamiltonians
\beq\label{ham}
 H^{\rm IV}=\sum_j \epsilon_j { n}_j+{V}^{\rm IV}\ \ {\rm and}
\ \ \   H^{\rm IS}=\sum_j \epsilon_j { n}_j+{ V}^{\rm IS},
\eeq
where ${ n}_j$ is the particle number operator and $\epsilon_j$ is the sp energy of the subshell $j$, respectively.

In the case of degenerate sp orbitals the Hamiltonians in  (\ref{ham}) can be described by the SO(8) model
and the eigenvalue problem can be solved analytically using group theory \cite{eng97,kot06}. 
If the sp orbitals do not degenerate the exact solution can be obtained only numerically 
using the Richardson-Gaudin equation \cite{duk04}. 
In the two-body case we do not apply these methods and 
we directly solve the eigenvalue problem of the Hamiltonians (\ref{ham}) using the full CI method. 

In the rest of this section we consider the sd shell, so there are 12 proton and 12 neutron modes, respectively.
The two-body matrix elements of the operators (\ref{isovecpot}) and (\ref{isoscapot}) 
can be found in \cite{Poves98} both for the $LS$ coupled and the $jj$ coupled sp bases. 
In this work we focus on the ground states, 
so we investigate two-nucleon systems, 
which are characterized with quantum numbers $J=0$ and $T=1$ in case of the IV pairing
and with quantum numbers  $J=1$ and $T=0$ in case of the IS pairing.

In the case of IV pairing the ground state are searched in the form  
\bea\label{isovecwf}
\Psi^{\rm IV}
=A_{\frac{1}{2},\frac{1}{2}}
\left[ c^\dagger_{s_{\frac{1}{2}}}\otimes c^\dagger_{s_{\frac{1}{2}}}\right]^{0,1}_{0,T_z}
+A_{\frac{3}{2},\frac{3}{2}}
\left[ c^\dagger_{d_{\frac{3}{2}}}\otimes c^\dagger_{d_{\frac{3}{2}}}\right]^{0,1}_{0,T_z}+
A_{\frac{5}{2},\frac{5}{2}}
\left[ c^\dagger_{d_{\frac{5}{2}}}\otimes c^\dagger_{d_{\frac{5}{2}}}\right]^{0,1}_{0,T_z} \nonumber \\
\eea  
and in the case of IS pairing the wave function is taken in the form
\bea\label{isoscawf}
\Psi^{\rm IS}=A_{\frac{1}{2},\frac{1}{2}}
\left[ c^\dagger_{s_{\frac{1}{2}}}\otimes c^\dagger_{s_{\frac{1}{2}}}\right]^{1,0}_{J_z,0}
+A_{\frac{3}{2},\frac{3}{2}}
\left[ c^\dagger_{d_{\frac{3}{2}}}\otimes c^\dagger_{d_{\frac{3}{2}}}\right]^{1,0}_{J_z,0}
+A_{\frac{5}{2},\frac{5}{2}}
\left[ c^\dagger_{d_{\frac{5}{2}}}\otimes c^\dagger_{d_{\frac{5}{2}}}\right]^{1,0}_{J_z,0}\nonumber\\
+A_{\frac{3}{2},\frac{5}{2}}
\left[ c^\dagger_{d_{\frac{3}{2}}}\otimes c^\dagger_{d_{\frac{5}{2}}}\right]^{1,0}_{J_z,0},
\eea  
where we apply the usual spectroscopic notation for the creation  tensor operators 
and we do not sign the  isospin value which is $\frac{1}{2}$. 
For a general state with angular momentum  $J=1$ the wave function 
$\left[ c^\dagger_{s_{\frac{1}{2}}}\otimes c^\dagger_{d_{\frac{3}{2}}}\right]^{1,0}_{J_z,0}$ 
should be in the superposition (\ref{isoscawf}), 
however we neglected this term since the used IS pairing in the ground state does not mix the state 
$\left[ c^\dagger_{s_{\frac{1}{2}}}\otimes c^\dagger_{d_{\frac{3}{2}}}\right]^{1,0}_{J_z,0}$ 
with the other components present in (\ref{isoscawf}) (see (\ref{tbmeisca})).

The $J=0$ and $T=1$ state (\ref{isovecwf}) can be described in Slater decomposed form 
according to the result of the section \ref{ciformsec} 
and we can use equation (\ref{onebodyentci}) to calculate the one-body entanglement entropy 
as the function of the parameters  $A^2_{\frac{1}{2},{\frac{1}{2}}}, A^2_{\frac{3}{2},{\frac{3}{2}}}$ and
$A^2_{\frac{5}{2},{\frac{5}{2}}}$. 

\begin{figure}[h]
\includegraphics[width=0.85\columnwidth]{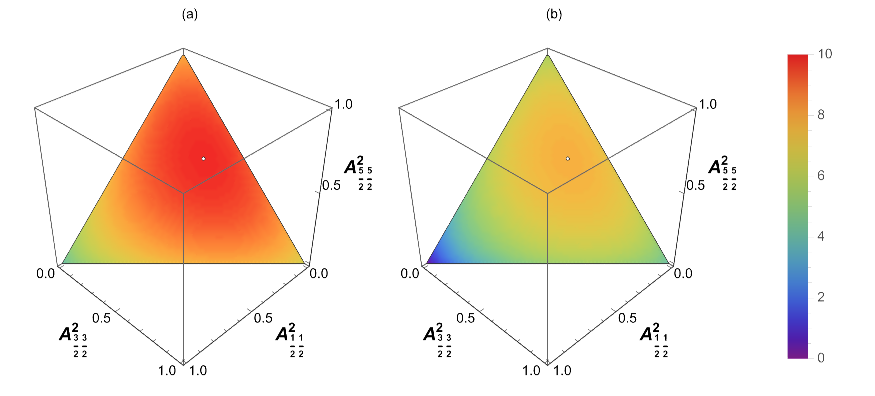}
\caption{\label{harom}
One-body entanglement entropy of $\Psi^{\rm IV}$ as function of the squared amplitudes. 
Part (a) shows the case $T_z=0$
and part (b) displays the case $T_z=1$. 
The small white circle shows the position of the maximum of the one-body entanglement entropy.}
\centering
\end{figure}

The results  are displayed in figure \ref{harom}. 
The values of the one-body entanglement entropies are coded using the colour scale given in figure \ref{harom}. 
In  part (a) and in part (b) of figure \ref{harom} 
results for the systems proton-neutron and neutron-neutron are shown. 
In both cases the maximal and minimal value of the one-body entanglement entropies occur at the same
position. The maximum of the one-body entanglement entropy for IV pairing, according to (\ref{maxent}), 
occurs where $A^2_{\frac{1}{2},{\frac{1}{2}}}=\frac{1}{6}, A^2_{\frac{3}{2},{\frac{3}{2}}}=\frac{1}{3}$ 
and $A^2_{\frac{5}{2},{\frac{5}{2}}}=\frac{1}{2}$. 
The actual maximal value of the one-body entanglement entropy depends on $T_z$. 
The value of the maximum is larger for  the proton-neutron case than the neutron-neutron systems, 
since two times more modes are present in the wave function with $T_z=0$ than in the case $T_z=1$.
The position of  the entropy minimum corresponds to the case when only one amplitude is different from zero 
and from table \ref{table2} it follows that 
the minimum is at $\left(A^2_{\frac{1}{2}},A^2_{\frac{3}{2}},A^2_{\frac{5}{2}}\right)=(1,0,0)$ for both cases. 
In the neutron-neutron case (part (b) of the figure \ref{harom}) the minimal value of the entropy is zero 
since this state is described only by one Slater determinant, in the proton-neutron case 
the minimum entropy is four (see table \ref{table2}) and the Slater rank of the corresponding state is two.

\begin{figure}[h]
\includegraphics[width=0.85\columnwidth]{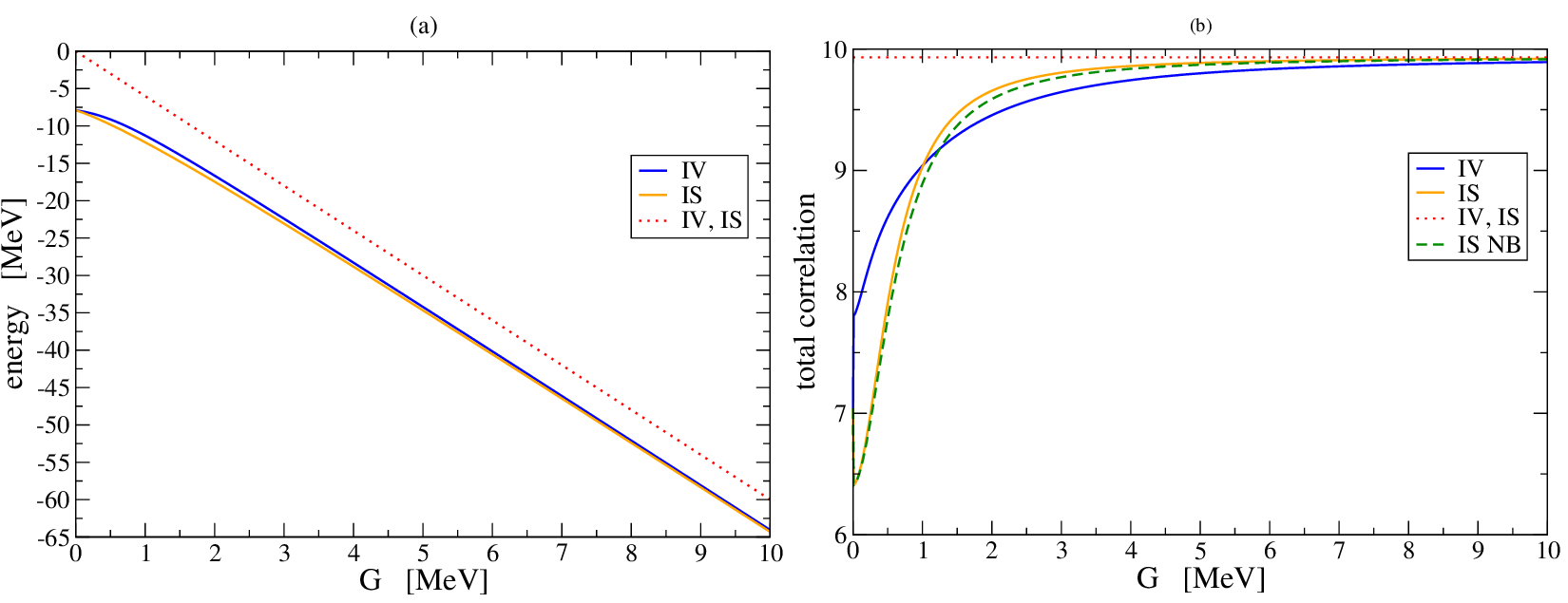}
\caption{\label{energy}
Energy (part (a)) and total correlation (part (b)) as the function of the strength of the interaction 
for the ground state with $J_z=0$ in case of proton-neutron system with IV and IS pairing interactions. 
The dotted lines correspond to degenerate sp orbitals. 
The abbreviation IS NB denotes calculations in natural basis in the case of IS pairing interaction.}
\centering
\end{figure}
Next we study how the ground state energy and the total correlation change as the function of the interaction strength 
if we use the Hamiltonians (\ref{ham}). 
We consider two cases. 	
First we put all the sp energies to zero and in the second case we apply nondegenerate sp energies.
In this work, every calculation that refers to nondegenerate sp energies applies the sp energies of the USD interaction \cite{Brown88}.
The results are displayed in figure \ref{energy}.  
In the case of the IV pairing, since in the ground state $J=0$, our basis is Slater basis, so it is natural basis 
and this means that the total correlation and the one-body entanglement entropy coincide.
In the case of IS pairing with nondegenerate sp energies 
the one-particle reduced density matrix of the ground state is nondiagonal. 
We have determined the total correlation in the natural basis to get the one-body entanglement entropy.
The figure \ref{energy} shows that the difference between the total correlation in the original basis and one-body entanglement entropy 
is not so large, this was observed also in \cite{Kruppa22}.

In the figure \ref{energy} the dotted lines indicate the case of degenerate sp levels, when the ground state wave function is independent from the interaction strength. Consequently, the one-body entanglement entropy is unaffected by the value of  $G$ or the considered type of pairing (IV or IS).
This property is explained in the \ref{app2}. 

More importantly, we will prove an interesting result in \ref{app2}: 
in the ground state the one-body entanglement entropy reaches the maximum value 
when the sp energies are degenerate and $J_z=0$ and $T_z=0$. 
This statement is true for both the IV  and IS pairings.
In these special cases, every natural orbital possesses the same occupation number, so
every orthonormal sp basis is natural basis. This means that the
occupation numbers and one-mode entropies are independent from the investigated orthonormal sp basis.

For nondegenerate sp energies the one-body entanglement entropy is influenced by the pairing strength. As pairing strength increases, the entropy approaches its maximum value.
We can observe an important trend the larger the binding energy 
the larger the one-body entanglement entropy in the cases of IV and IS pairing.

\begin{figure}[h]
\includegraphics[width=0.98\columnwidth]{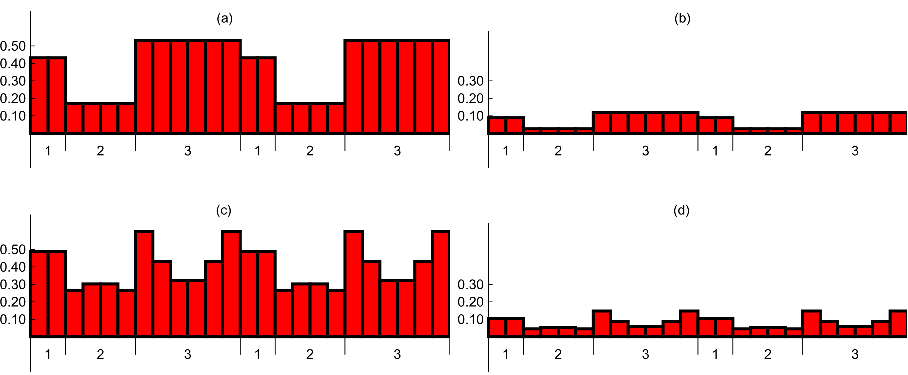}
\caption{\label{occfig}
One-mode entropies (parts (a) and (c)) and occupation numbers (parts (b) and (d)) of the ground state 
for the proton-neutron system with $J_z=0$. 
The results for IV pairing are in parts (a) and (b) the results for IS pairing are in parts (c) and (d). For the notation of the labelling of the modes see the text.}
\centering
\end{figure}

Next we investigate the one-mode entropies. We use the following order of the sp orbitals. 
First come all sp orbitals which has isospin projection $-\frac{1}{2}$ 
and after this come all sp orbitals with isospin projection $\frac{1}{2}$. 
For a fixed isospin projection the modes are arranged in the order 
$1s_{\frac{1}{2}}, 0d_{\frac{3}{2}}$ and $0d_{\frac{5}{2}}$, 
in the figures we use the notation $1,\ 2$ and $3$ for these orbitals. 
The modes with a given $n,l,j,t_z$ quantum numbers are arranged 
in increasing order of their $j_z$ quantum numbers. 

Interestingly, for the $J_z=T_z=0$ ground state in the case of degenerate sp energies 
all the modes have the same occupation numbers and consequently the one-mode entropies are identical. 
This statement is true both for IV and IS cases (see \ref{app2}). 
In figure \ref{occfig} we display the one-mode entropies  and occupation numbers when the sp energies are nondegenerate.  
In calculations with fixed $G$ value the strength of the interactions are always $G=2$ MeV.

The use of nondegenerate orbitals causes that occupation numbers depend on the $j$ values. 
In the case of $J=0$ (see parts (a) and (b) of figure \ref{occfig}) 
the occupation numbers (one-mode entropies) does not depend on the $j_z$ values of the mode. 
However, if we consider the 
IS pairing ($J=1$, see part (c) and (d) of the figure \ref{occfig}), 
the one-mode entropies depend on the value of $j_z$. 
The values of one-mode entropies are closely follows the pattern of the occupation numbers. 
This correlation was observed also in \cite{PerezObiol23a}. 
The proton and neutron one-mode entropies are identical 
since in the considered case $T_z=0$ and the interaction is isospin conserving.
The IS pairing case part (c) of figure \ref{occfig} strikingly demonstrate the former general findings: 
if $J_z=0$, the modes $(n,l,j,j_z,t_z)$ and $(n,l,j,-j_z,t_z)$ have identical one mode entropies,
if $T_z=0$ and the isospin is fixed  the modes  $(n,l,j,j_z,t_z)$ and $(n,l,j,j_z,-t_z)$
also have the same one-mode entropies.
So in the figure every values of the one-mode entropies 
appear at least four times due to the modes  $( n,l,j,\pm j_z, \pm t_z)$.

\begin{figure}[h]
\includegraphics[width=0.98\columnwidth]{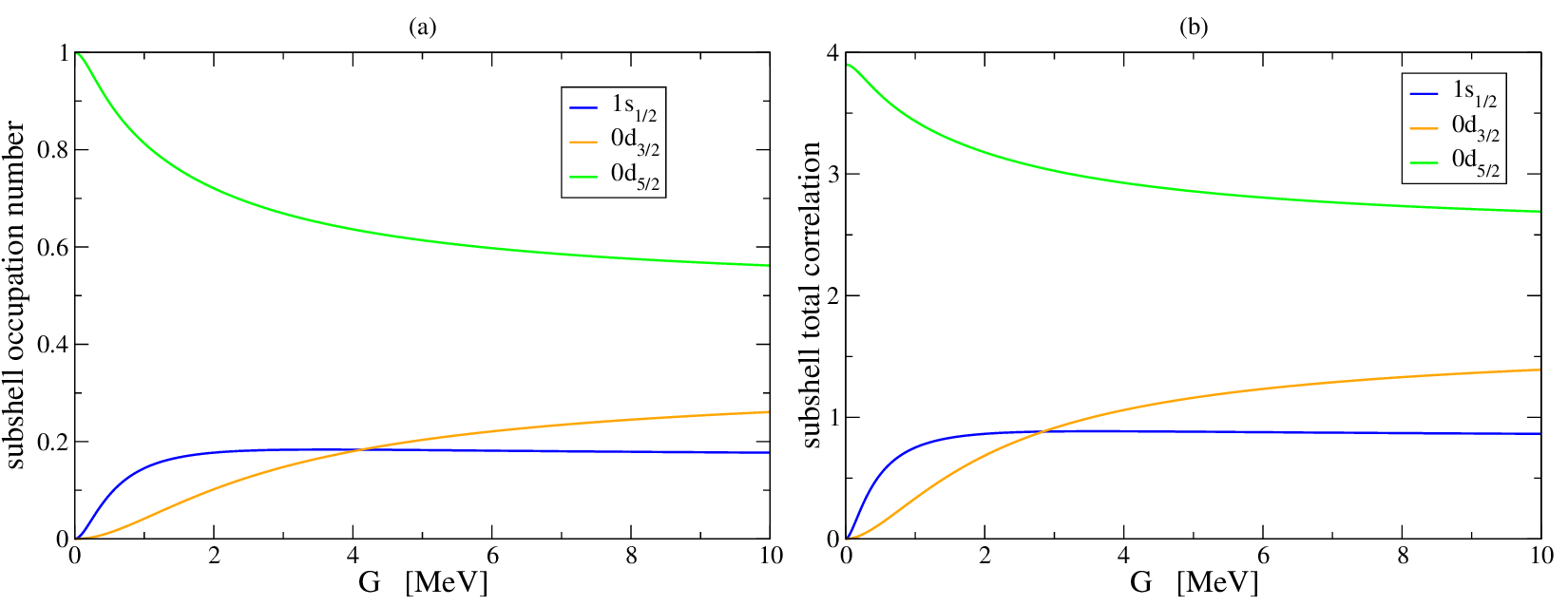}
\caption{\label{subshell}
The proton subshell occupation numbers (part (a)) and the proton subshell total correlation (part (b)) 
as the function of the interaction strength. 
The proton-neutron system is considered in the ground state with IV pairing.}
\centering
\end{figure}

Similarly to the subshell occupation numbers   
$\underset{j_z}{\sum} \langle c^\dagger_{ n,l,j,j_z,t_z}c_{ n,l,j,j_z,t_z}\rangle$ 
we can define the subshell total correlation $\underset{j_z}{\sum} S^{\rm 1M}_{ n,l,j,j_z,t_z}$.
The dependence of these two quantities on the interaction strength is displayed in figure \ref{subshell} 
for the $T_z=0$ ground state of the IV pairing.
For our sp energies we have the following relation 
$\epsilon_{0d_{\frac{5}{2}}}<\epsilon_{1s_{\frac{1}{2}}}<\epsilon_{0d_{\frac{3}{2}}}$.
For small interaction strength one can observe the following pattern, 
the smaller the sp energy of the subshell the larger the subshell total correlation. 
This trend however is not valid for larger values of the interaction strength $G$. 
Generally, we can say that the subshell total correlation follows the trend of the subshell occupation numbers 
except a narrow region of the interaction strength roughly between $G=3$ MeV and $G=4$ MeV.

\begin{figure}[h]
\includegraphics[width=0.85\columnwidth]{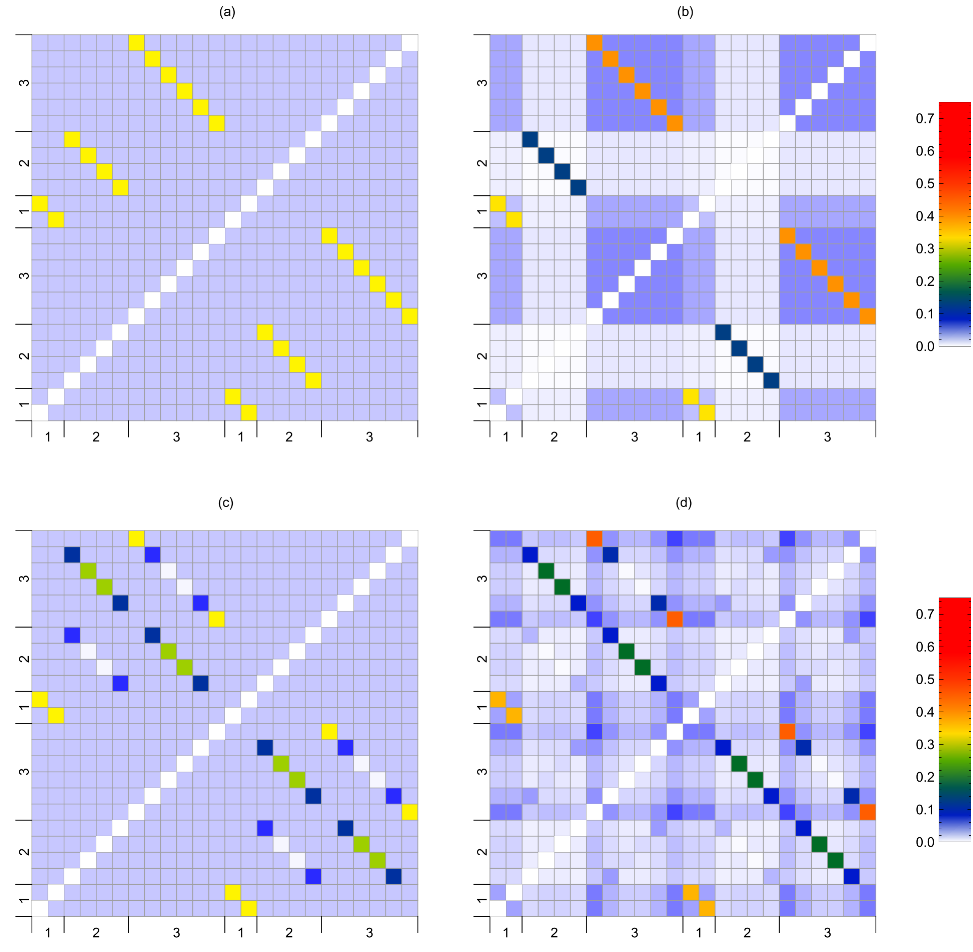}
\caption{\label{miffig}
The mutual informations of mode pairs for the $J_z=0$ ground state of proton-neutron system 
for IV pairing (parts (a) and (b)) and for IS pairing (parts (c) and (d)). 
The parts (a) and (c) display the results when the sp energies are degenerate, in parts (b) and (d) the sp energies are nondegenerate. For the notation of the labelling of the modes see the text.}
\end{figure}

The calculated mutual informations for the ground states is displayed in figure \ref{miffig} for both IV and IS pairings.
We considered the cases when the sp energies are degenerate and nondegenerate.  
The arrangement of the modes is the same as described in the discussion of the one-mode entropies. 
The values of the mutual informations are coded using the colour scale given in figure \ref{miffig}.  
The diagonal squares, representing cases where the row and column modes are the same, are displayed in white in the figures, as mutual information is defined only between different modes.

The simplest pattern of mutual informations is in part (a) of figure \ref{miffig} (IV pairing and  zero sp energies), where we can see that there are only two different values of the mutual informations.
If we substitute the amplitudes (\ref{solivec}) of the wave function into table \ref{tablemi},
we get that the mutual informations of all associated mode pairs
are $h(s)$ and the mutual informations of the nonassociated pairs are $2H(1-s)-H(1-2s)$, 
where $s=\left({\underset{j}{\sum} (2j+1)}\right)^{-1}$.
The associated modes appear in  angular momentum coupled two-nucleon components of the total wave function. 
In the case of IV pairing there are three such components in the wave function (\ref{isovecwf}),  
these correspond to strikingly coloured squares along three slanted lines in part (a) and (b) of  figure \ref{miffig}. 
In the case of IS pairing there are four angular momentum coupled components in wave function (\ref{isoscawf}), 
so we can notice in part (c) of figure \ref{miffig} 
high-contrasted coloured squares along four slanted lines.

The simple pattern of part (a) of figure \ref{miffig} is modified if the sp energies are nondegenerate 
(part (b) of figure \ref{miffig}). 
We can see that in the case of IV pairing with nondegenerate sp energies 
there are nine different values of the mutual informations. 
This follows form the expressions of table \ref{tablemi}.
The associated modes of the orbital $d_{\frac{5}{2}}$ have the largest mutual information 
and the second largest one is the associated modes of the orbital $s_{\frac{1}{2}}$.
The observed order of the values of mutual informations corresponds to the shell structure of the sp energies 
$\epsilon_{0d_{\frac{5}{2}}}<\epsilon_{1s_{\frac{1}{2}}}<\epsilon_{0d_{\frac{3}{2}}}$.

In the case of the IS pairing the situation is more complicated. 
The four different angular momentum couplings  (see \ref{isoscawf}) can be noticed 
if we view part (c) of the figure \ref{miffig}. 
However, if the sp energies are not degenerate (part (d) of the figure \ref{miffig}) the mutual informations of the mode pairs coming from the coupling $\left[ c^\dagger_{d_{\frac{3}{2}}}\otimes c^\dagger_{d_{\frac{3}{2}}}\right]^{1,0}_{J_z,0}$ are very small, which also reflects the structure of the sd shell. The strength of the influence of the shell structure on the mutual informations of course depends on the used interaction. 

In the \ref{app3} for Slater basis we will prove that fixing a mode $i$ and considering all the mode pairs $(i,j)$ 
the associated mode pair $(i,i_{\rm ass})$ has the largest value of the mutual information, where
 $i_{\rm ass}$ denotes the index of the mode that is associated to mode $i$.
This means that in the parts (a) and (b)  of figure \ref{miffig} if we select a column 
then in this column the associated mode pair has the largest mutual information value. 

We have seen that for our interactions in the degenerate sp energy case 
the total correlation of the $J_z=T_z=0$ ground state is the same for IV and IS pairings.
One can compare part (a) and part (c) of figure \ref{miffig} 
and notice that the mutual informations differentiate between the wave functions (\ref{isovecwf}) and (\ref{isoscawf}). 
The mutual informations reflects the correlations in the wave function 
better than the one-mode entropies (or total correlation) 
in the $jj$ coupled basis of the sd shell (see later the result of the $LS$ coupled basis).

\begin{figure}[h]
\includegraphics[width=0.85\columnwidth]{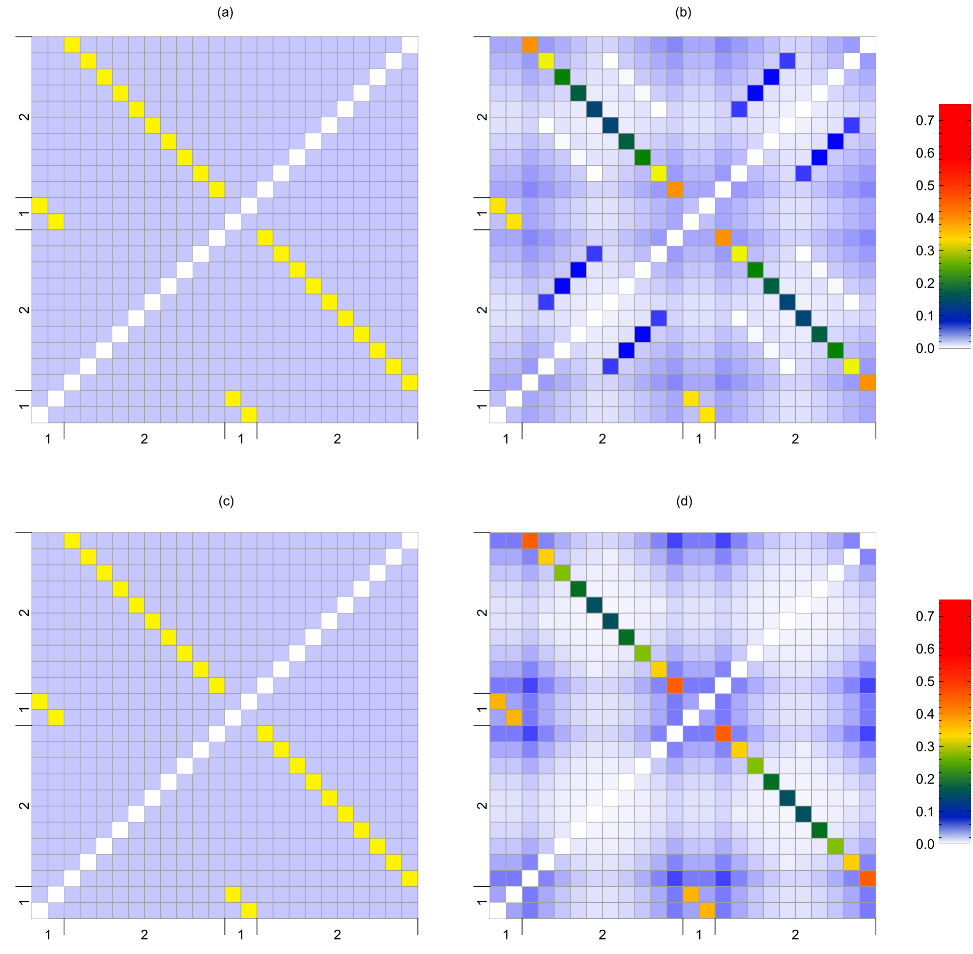}
\caption{\label{lsfig}
The mutual informations of mode pairs for the $J_z=0$ ground state of the proton-neutron system 
for IV pairing (parts (a) and (b)) and for IS pairing (parts (c) and (d)). 
The parts (a) and (c) display the results when the sp energies are degenerate. 
The modes are $LS$ coupled orbitals.
For the notation of the labelling of the modes see the text.}
\end{figure}

If the analysed case is Slater basis
 (that is parts (a) and (b) of figure \ref{miffig} and parts (a) and (c) of figure \ref{lsfig}) it is worth to compare the associated mode pairs and nonassociated mode pairs also from the viewpoint of two-mode entropies. In these cases we observed that the two-mode entropies related to nonassociated mode pairs are greater than those connected to associated mode pairs.

Beyond the usual symmetries of spherical scalar and isospin scalar interactions,
the IV and IS pairing interactions possess further symmetries, since in these cases the total orbital 
angular momentum $L$, the total spin $S$ and their projections ($L_z$ and $S_z$) are also good 
quantum numbers. 
These additional symmetries suggest that the $LS$ coupled sp orbitals may be better suited for 
the considered  Hamiltonians than the $jj$ coupled ones, so we also determined the mutual informations in $LS$ 
coupled sp basis and displayed the results in figure \ref{lsfig}.

This figure applies the following arrangement for the 
$a^\dagger_{n,l,m,\frac{1}{2},s_z,\frac{1}{2},t_z}\vert 0 \rangle$ modes of the basis. 
Similarly to  figure \ref{miffig}, in the applied order of the modes first we consider the 12 modes 
that have isospin projection $-\frac{1}{2}$ and after these 
we consider the other 12 modes with isospin projection $\frac{1}{2}$. 
The modes with a given isospin projection are arranged so that first the two modes of the $1s$ orbital are enumerated, 
then the 10 modes of the $0d$ orbital are enumerated. 
In the figure we use the notation $1$ and $2$ for these orbitals. 
The modes with given $n,l,t_z$ quantum numbers are grouped according to their their $s_z$ quantum number 
and arranged so that first the modes with $s_z=-\frac{1}{2}$, 
then the modes with $s_z=\frac{1}{2}$ are listed.
The modes with given $n,l,t_z,s_z$ quantum numbers are arranged in increasing order of their $m$ quantum numbers. 
Similar to figure \ref{miffig}, this order of the modes determines 
the arrangement according to the bottom-top and the left-right direction of the figure.

Before the discussion of figure \ref{lsfig}, we note that the applied $LS$ coupled sp basis 
and the former $jj$ coupled sp basis have a few common
elements, namely the modes with $l=0$ or $\vert j_z \vert =\frac{5}{2}$.
So for mutual informations between such modes the same value can be read out 
both from figure \ref{miffig} and from figure \ref{lsfig}.
For example, red squares in part (d) of figure \ref{miffig} and  
in part (d) of figure \ref{lsfig}
refer to such mode pairs.
 
For the wave functions analysed on the figures the common modes are also elements of the Slater bases,
even in the case of nondegenerate sp energies.
This peculiarity follows from the facts, that
the modes with different $l$ are not coupled, furthermore $J_z$ and $T_z$ are good quantum numbers.

If the sp energies are degenerated (parts (a) and (c) in figure \ref{lsfig}), a given mode of the 
 $LS$ coupled basis can compose a Slater determinant with only one unique mode, since
the applied pairing interactions do not couple modes with different $l$ quantum numbers
and conserve the quantum numbers $L_z$, $S_z$ and $T_z$.
This observation implies that the $LS$ coupled sp basis is Slater basis for the degenerate cases.
However, this basis is not Slater basis in the nondegenerate cases  (parts  (b) and (d) in figure \ref{lsfig}), 
since then the spin-orbit coupling, 
which is taken into account by the difference between the sp energies 
for $j=\frac{3}{2}$ and $j=\frac{5}{2}$, 
destroys the symmetries for projections of orbital angular momentum and spin. 

If the sp energies are degenerate, the same occupation number belongs to every mode 
(see \ref{app2}), so there are 
only two possibilities for the value of the mutual informations,
according to that the studied two modes are associated or not.
This results the simple structure of parts  (a) and (c).
However parts (a) and (c) refer to wave functions with different  
quantum numbers, the difference occurs  only in the signs of their Slater determinants and
the values of the coefficients $\lambda$ are the same in both cases,
so the values of mutual informations are also the same.
This leads to the interesting situation that mutual informations in the Slater basis (which modes are 
natural orbitals) do not distinguish these wave functions,
while mutual informations of the former $jj$ coupled sp basis reveal the differenties between the two wave 
functions.

If we compare the results for the $J=0$ case with degenerate sp energies in the two different sp bases
(part (a) of figure \ref{miffig} and part (a) of figure \ref{lsfig}), we can observe
that the two possible values of the mutual informations are the same.
For this wave function both sp bases are Slater basis and 
the possible values of the $\lambda$ coefficients of the Slater decomposition 
are independent from the chosen Slater basis \cite{Chen15}. 

Finally, we note that when mutual informations are studied in $LS$ coupled basis an interesting symmetry occurs, which
is also valid in the nondegenerate cases. 
In figure \ref{lsfig} the 10$\times 10$ subarrays, those refer to modes with $l=2$, 
have a diagonal symmetry axis in the top left to bottom right direction. 
This means that for the wave functions of the figure
the $I_{(n,l,m,s_z,t_z;n,l,m',s'_z,t'_z)}=I_{(n,l,-m',-s'_z,t_z;n,l,-m,-s_z,t'_z)}$ 
symmetry is fulfilled for the mutual informations, where the 
$I_{(n_i,l_i,m_i,{s_z}_i,{t_z}_i;n_j,l_j,m_j,{s_z}_j,{t_z}_j)}\equiv I_{(i,j)}$
notation is applied.

\section{Summary}
\label{section8}

We examined angular momentum and isospin coupled single configuration two-nucleon states in Slater decomposed form and obtained analytical results for the one-mode entropies. We studied a specific form of configuration interaction type scalar wave function and derived analytic expressions for one-mode entropies and mutual informations as functions of the amplitudes. 

In the presence of interaction we focused on isovector and isoscalar pairing force and the numerical calculations are carried out in the sd shell. Initially, we utilized degenerate single-particle orbitals, allowing us to derive analytical results using the configuration interaction method.
We proved analytically that, for both isovector and isoscalar pairing interactions, the ground state of the proton-neutron system shows maximal one-body entanglement entropy and total correlation when single-particle orbitals are degenerate and total angular momentum projection is zero.
Subsequently, we investigated realistic scenarios where the single-particle orbitals were not degenerate.
We demonstrated the natural expectation that the subshell total correlation
almost always follows the pattern of the change of occupation numbers if the interaction strength is changed.

The mutual information for all pairs of modes is presented on a square grid. For proton-neutron system in $jj$ coupled single particle basis, the observed patterns of the mutual informations of the  ground state of the IV interaction can be explained by the associated modes and the form of the angular momentum couplings. In the case of degenerate single-particle orbitals, the pattern is clear for both types of pairing interactions. In a realistic scenario, when single-particle orbitals are nondegenerate, the main patterns remain mostly unchanged with slight variations. The influence of the shell structure (the order of the energies of the single-particle orbitals) on the values of mutual informations can be observed.

The quantities of one-mode entropies, mutual informations, and total correlation are known to be dependent on the basis. For the isovector and isoscalar pairing interactions, the total orbital momentum, spin and their projections are also good quantum numbers in the case of degenerate single-particle energies. It is expected that for the isoscalar case the observed patterns of the entanglement measures are much simpler when the $LS$ orbitals are considered for the modes. We also calculated the mutual informations for the $LS$ coupled modes and observed that the patterns  for the isoscalar cases are indeed simplified.

Regardless of the number of particles involved, we used the $jj$ coupling scheme to derive some conclusions about rotational symmetries. If the total angular momentum projection is zero, then the mode entropies do not depend on the sign of the single-particle angular momentum projection. Similarly, for systems conserving isospin, if the total isospin projection is zero, then the proton and neutron modes with identical quantum numbers have the same one-mode entropies. This can be used to measure the degree of isospin breaking.

In the near future, we intend to study entanglement measures in the presence of isovector and isoscalar pairing interactions involving more than two particles.
\ack
This work was supported by the
National Research, Development and Innovation Fund of Hungary,
financed under the K18 funding scheme with project nos. K 128729
and K 134983. \"O.L. acknowledges financial support
by the Hans Fischer Senior Fellowship programme funded by the Technical University
of Munich – Institute for Advanced Study and by the Quantum Information National Laboratory
of Hungary and by the Center for Scalable and Predictive methods
for Excitation and Correlated phenomena (SPEC),
funded as part of the Computational Chemical Sciences, FWP 70942,
by the U.S. Department of Energy (DOE), Office of Science, Office of Basic Energy Sciences, Division of Chemical Sciences, Geosciences, and Biosciences at Pacific Northwest National Laboratory.
\appendix 
\section {\label{app1}  Isovector single configurations and spherical scalar single configurations}

In this section we give some further formulas for single configurations. 

The $T_z=0$ isovector single configurations can be written as
\bea
\psi^{J,J_z,T=1,T_z=0}_{ \rho,j, \rho,j}=
\underset{j_z \in {\cal S}_{j,j,J_z}}{\sum}
C^{J,J_z}_{j,j_z,j,\overline{j_z}}
c_{ \rho,j,j_z,-\frac{1}{2}}^\dagger c_{ \rho,j,\overline{j_z},\frac{1}{2}}^\dagger|0\rangle
\eea
 for same subshells with even $J$
and
\bea
\psi^{J,J_z,T=1,T_z=0}_{ \rho_1,j_1, \rho_2,j_2}&=&
\frac{1}{\sqrt{2}} 
\underset{{j_z}\in{\cal S}_{j_1,j_2,J_z}}{\sum}
C^{J,J_z}_{j_1,j_z,j_2,\overline{j_z}}
\bigg[ c_{ \rho_1,j_1,j_z,-\frac{1}{2}}^\dagger c_{ \rho_2,j_2,\overline{j_z},\frac{1}{2}}^\dagger|0\rangle
 \nonumber \\
&&+ c_{ \rho_1,j_1,j_z,\frac{1}{2}}^\dagger c_{ \rho_2,j_2,\overline{j_z},-\frac{1}{2}}^\dagger|0\rangle\bigg] 
\eea
 for different subshells with arbitrary $J$.

The $T_z=\pm 1$ isovector single configurations can be written as
\bea
\psi^{J,J_z,T=1,T_z=\pm 1}_{ \rho,j, \rho,j}=
\sqrt{2}
\underset{j_z \in {\cal S}_{j,j,J_z},j_z<\frac{J_z}{2}}
{\sum}
C^{J,J_z}_{j,j_z,j,\overline{j_z}}
c_{ \rho,j,j_z,\pm\frac{1}{2}}^\dagger c_{ \rho,j,\overline{j_z},\pm\frac{1}{2}}^\dagger|0\rangle 
\eea
for the same subshells with even $J$ and
\bea
\psi^{J,J_z,T=1,T_z=\pm 1}_{ \rho_1,j_1, \rho_2,j_2}= 
\underset{{j_z}\in{\cal S}_{j_1,j_2,J_z}}
{\sum}
C^{J,J_z}_{j_1,j_z,j_2,\overline{j_z}}
c_{ \rho_1,j_1,j_z,\pm\frac{1}{2}}^\dagger c_{ \rho_2,j_2,\overline{j_z},\pm\frac{1}{2}}^\dagger|0\rangle
\eea
for different subshells with arbitrary $J$.

For spherical scalar single configurations in Slater basis we can obtain 
simple expressions for more quantities both in isovector and in isoscalar cases.
These analytical results are summarized in tables \ref{table2} and \ref{table3}.
We note that in these special cases the  $\lambda$ coefficients of different associated
mode pairs are the same, so
every mode of the Slater basis that present in the wave functions has the same one-mode entropy, 
$h(4\lambda^2)$.

\begin{table}[H]
\caption{\label{table2} Characteristics of the angular momentum and isospin coupled single configurations 
for $J=0$ state. 
The Slater rank, occupation number $4\lambda^{2}$ and the one-body entanglement entropy are given. 
For same subshells, due to antisymmetrization there is no $T=0$ state.}
\begin{indented}
\item[]\begin{tabular}{cccccc}
\br
T&$T_z$ & subshells &   Slater rank &  $4\lambda^2$ & $S^{{\rm 1B}}$ \\ 
\mr
1&$\pm 1$ & same   &   $\frac{2j+1}{2}$
& $\frac{2}{2j+1}$  &  $(2j+1){\rm h}\left (\frac{2}{2j+1}\right)$ \\
1&$\pm 1$  & different  &    $2j+1$
& $\frac{1}{2j+1}$ &  $2(2j+1){\rm h}\left (\frac{1}{2j+1}\right)$  \\
1&0 & same  &   $2j+1$
& $\frac{1}{2j+1}$  & $2(2j+1){\rm h}\left (\frac{1}{2j+1}\right)$    \\
1&0 & different  &  $4j+2$
&  $\frac{1}{2}\frac{1}{2j+1}$ & $2(4j+2){\rm h}\left (\frac{1}{2}\frac{1}{2j+1}\right)$  \\
0&0 & different  &   $4j+2$
& $\frac{1}{2}\frac{1}{2j+1}$ & $2(4j+2){\rm h}\left (\frac{1}{2}\frac{1}{2j+1}\right)$   \\
\br
\end{tabular}
\end{indented}
\end{table}

\begin{table}[H]
\caption{\label{table3} The mutual informations of angular momentum and isospin coupled single 
configurations in the case of $J=0$ in Slater basis. 
In the  case of nonassociated modes we assume that both modes present in the wave function,
otherwise the mutual information is zero.
For same subshells, due to antisymmetrization there is no $T=0$ state.}
\begin{indented}
\item[]\begin{tabular}{ccccc}
\br
T&$T_z$  &  subshells & the  mode pair  & mutual information    \\ 
\mr
1&$\pm 1$ & same   &   associated  & ${\rm h}\left (\frac{2}{2j+1}\right)$ \\
1&$\pm 1$ & different  &   associated  & ${\rm h}\left (\frac{1}{2j+1}\right)$  \\
1&0 & same   &  associated &  ${\rm h}\left (\frac{1}{2j+1}\right)$ \\
1&0 & different  &   associated & ${\rm h}\left (\frac{1}{2}\frac{1}{2j+1}\right)$    \\
0&0 & different  &   associated & ${\rm h}\left (\frac{1}{2}\frac{1}{2j+1}\right)$ \\
1&$\pm 1$  & same   & not  associated  
&  $2{\rm H} \left(1-\frac{2}{2j+1}\right)-{\rm H} \left(1-\frac{4}{2j+1}\right)$\\
1&$\pm 1$  & different  &  not  associated  
& $2{\rm H} \left(1-\frac{1}{2j+1}\right)-{\rm H} \left(1-\frac{2}{2j+1}\right)$  \\
1&0 & same  & not  associated 
& $2{\rm H} \left(1-\frac{1}{2j+1}\right)-{\rm H} \left(1-\frac{2}{2j+1}\right)$\\
1&0 & different  & not  associated 
& $2{\rm H} \left(1-\frac{1}{2}\frac{1}{2j+1}\right)-{\rm H} \left(1-\frac{1}{2j+1}\right)$  \\
0&0 & different  & not  associated 
& $2{\rm H} \left(1-\frac{1}{2}\frac{1}{2j+1}\right)-{\rm H} \left(1-\frac{1}{2j+1}\right)$ \\
\br
\end{tabular}
\end{indented}
\end{table}
\section{\label{app2} Exact solutions of two-body pairing Hamiltonians}

If the studied Hamiltonians (\ref{ham}) do not contain the terms of sp energies, 
we can obtain simple analytical solutions 
for the ground states.
For the sake of simplicity, in this section we restrict us to one main shell. 

We start the derivation of the solutions by the formulas of \cite{Poves98} 
for the two-body matrix elements of the applied IV and IS pairing interactions.
In the $jj$ coupled basis these matrix elements are 
\bea\label{tbmeivec}
\langle a,b,J,T\vert  V^{\rm IV} \vert c,d,J,T\rangle 
&=&
-G\sqrt{\left(j_a+\frac{1}{2}\right)\left(j_c+\frac{1}{2}\right)}  \delta_{a,b}\delta_{c,d}\delta_{J,0}\delta_{T,1},
\eea
for the IV pairing and 
\bea\label{tbmeisca}
\langle a,b,J,T  \vert V^{\rm IS} \vert c,d,J,T  \rangle 
&=&-G\frac{2(-1)^{j_a-j_c}}{\sqrt{1+\delta_{a,b}}\sqrt{1+\delta_{c,d}}} \nonumber \\
&\times& \sqrt{(2j_a+1)(2j_b+1)(2j_c+1)(2j_d+1)} \nonumber \\
&\times& 
\left\{
\begin{array}{ccc}
\frac{1}{2} & j_a & l_a \\
j_b & \frac{1}{2} & 1
\end{array}\right\}
\left\{
\begin{array}{ccc}
\frac{1}{2} & j_c & l_c \\
j_d & \frac{1}{2} & 1
\end{array} \right\}\nonumber \\
&\times& 
\delta_{A,B}\delta_{C,D}\delta_{J,1}\delta_{T,0}
\eea
for the IS case,
where the triplets $\{ n,l,j\}$ are denoted by lower case letters, 
the pairs $\{ n,l\}$ are denoted by capital letters and we changed the notation of \cite{Poves98} 
for the coupling of the interaction. 
In our notation $G>0$ refers for the cases when the energy of the ground state of $V^{\rm IV}$
or $V^{\rm IS}$ is negative.

The ground state of $V^{\rm IV}$ with fixed $T_z$ can be written in CI form as
\bea
\Psi^{\rm IV}(T_z)=\underset{i}{\sum} A_i\Psi^{\rm IV}_i(T_z),
\eea
where the single configurations are denoted as
\bea
\Psi^{\rm IV}_i(T_z)
=\frac{1}{\sqrt{2}}\left[c^\dagger_{a{(i)}}\otimes c^\dagger_{a{(i)}}\right]^{J=0,T=1}_{J_z=0,T_z},
\eea
where the $a{(i)}\equiv( n{(i)},l{(i)},j{(i)})$ notation is applied for the triplets of quantum numbers.
The notation refers that the given quantum numbers are determined by the index $i$.

The variational method gives a matrix eigenvalue problem  $\underline V^{\rm IV}{\bf A}=E {\bf A}$, 
where the column vector $\bf A$ has components $A_i$ 
and the elements of Hamiltonian matrix $\underline V^{\rm IV}$ is given by 
$\underline V^{\rm IV}_{i,j}=\langle\Psi^{\rm IV}_i(T_z)\vert  V^{\rm IV}\vert\Psi^{\rm IV}_j(T_z)\rangle$, 
where the right hand side is just a two-body matrix element. As see from (\ref{tbmeivec})
this value is independent from $T_z$. 

One can notice that the Hamiltonian matrix can be turned into the form 
\bea\label{dyadic1}
{\underline V^{\rm IV}}
=-\frac{G}{2}{\bf v} {\bf v}^{\rm t},
\eea
 where the upper index t denotes the transposed, 
and the column vector $\bf v$ has the components 
\beq
v_i=\sqrt {2j{(i)}+1}.
\eeq
The right hand  side of (\ref{dyadic1}) contains 
the dyadic product ${\bf v}\otimes{\bf v}\equiv{\bf v} {\bf v}^{\rm t}$. 

It is known \cite{Dattoro19}, that for the dyad ${\bf v}\otimes{\bf v}$ an eigenvalue is ${\bf v}^t {\bf v}$ 
with the eigenvector $\bf v$, while the other eigenvalues are zero.
This means that the ground state energy in the case of IV pairing is
\beq\label{eivec}
-\frac{G}{2}\sum_i (2 j{(i)}+1)
\eeq
and coefficients $A_i$ can be chosen as the components of the normalized eigenvector $\bf v$.
For this choice
\beq\label{solivec}
A_i=\frac{\sqrt{2j{(i)}+1}}{\sqrt{\underset{i}{\sum} (2 j{(i)}+1})}.
\eeq
If we use (\ref{cimode1}) we get for the one-mode entropies
\beq
S_k^{1M}(T_z)=
\left\{
\begin{array}{ll}
{\rm h}\left(\frac{1}{\underset{i}{\sum} (2 j{(i)}+1)}\right)& {\rm for}\ T_z=0,\\
{\rm h}\left(\frac{2}{\underset{i}{\sum}(2 j{(i)}+1)}\right)& {\rm for}\ T_z=\pm 1,
\end{array}
\right.
\eeq 
and from it follows that each mode has the same one-mode entropy. 
For $T_z=0$ the occupation numbers are of the form that the particle number is divided by the number of modes 
present in the wave function, so the one-body entanglement entropy is maximal \cite{Kruppa21}.
For $T_z=\pm 1$ cases this statement true only if we restrict ourselves to the modes with fixed isospin projections.

In order to prove that in the case of the IS pairing the one-body entanglement entropy takes its maximal value for the ground 
state with quantum numbers $J=1$ and $J_z=0$ we switch to the $LS$ coupling scheme. 
Here we abbreviate the full form of the mode creation operators  
$a^\dagger_{n,l,m,\frac{1}{2},s_z,\frac{1}{2},t_z}$
as $a^\dagger_{n,l,m,s_z,t_z}$ and we use uncoupled $LS$ formalism. 
In this basis the two-nucleon Slater determinants can be written as 
\beq\label{tbmeiscawf}
\Psi_{\alpha,\beta}=a^{\dagger}_{n_\alpha,l_\alpha,{m}_\alpha,{s_z}_\alpha,{t_z}_\alpha}
a^{\dagger}_{n_\beta,l_\beta,{m}_\beta{s_z}_\beta,{t_z}_\beta}|0\rangle.
\eeq
From the $jj$ coupled matrix elements (\ref{tbmeisca}) 
one can show that the matrix elements of the
 uncoupled $LS$ formalism have the form 
\bea\label{isoscaLS}
&&\langle \Psi_{\alpha,\beta}\vert  V^{\rm IS}
\vert \Psi_{\gamma,\delta}\rangle  
\nonumber \\
&&=-G(-1)^{l_\alpha-{m}_\alpha}(-1)^{\frac{1}{2}-{t_z}_\alpha}
 (-1)^{l_\gamma-{m}_\gamma}(-1)^{\frac{1}{2}-{t_z}_\gamma}
\nonumber \\
&&
\times \delta_{n_{\alpha},n_{\beta}}\delta_{l_{\alpha},l_{\beta}}\delta_{{m}_\alpha+{m}_\beta,0}
\delta_{{t_z}_\alpha+{t_z}_\beta,0}
\delta_{n_{\gamma},n_{\delta}}\delta_{l_{\gamma},l_{\delta}}\delta_{{m}_\gamma+{m}_\delta,0}
\delta_{{t_z}_\gamma+{t_z}_\delta,0} \nonumber \\
&& \times \left\{
\begin{array}{ll}
\frac{1}{2}&  {\rm if } {s_z}_\alpha+{s_z}_\beta={s_z}_\gamma+{s_z}_\delta=0, \\
1& {\rm if } {s_z}_\alpha+{s_z}_\beta={s_z}_\gamma+{s_z}_\delta=\pm 1, \\
0& {\rm if } {s_z}_\alpha+{s_z}_\beta\neq{s_z}_\gamma+{s_z}_\delta.
\end{array}
\right.
\eea

Now we can perform a procedure that is similar to the previous one. 
First we introduce the 
\bea
\Psi^{\rm IS}_i(S_z)=a^{\dagger}_{n{(i)},l{(i)},{{{m}{(i)}}},{{s_z}{(i)}},{{t_z}{(i)}}}
a^{\dagger}_{n{(i)},l{(i)},-{m}{(i)},S_z-{s_z}{(i)},-{t_z}{(i)}}|0\rangle
\eea
notation, then we write the ground state of $V^{\rm IS}$ with fixed $S_z$  as
\bea\label{ciis}
\Psi^{\rm IS}(S_z)&=&
\underset{(n,l,{m},{s_z},{t_z})<(n,l,-{m},S_z-s_z,-{t_z})}{\sum{}} 
B_{n,l,{m},{s_z},{t_z}}
a^{\dagger}_{n,l,{m},{s_z},{t_z}}
a^{\dagger}_{n,l,-{m},S_z-{s_z},-{t_z}}|0\rangle \nonumber \\
&\equiv&\underset{i}{\sum} B_{i}
\Psi^{\rm IS}_i(S_z),
\eea
where in the first line the summation runs over the 5-tuples  $(n,l,{m},{s_z},t_z)$ and
the $(n,l,{m},{s_z},{t_z})<(n,l,-{m},S_z-s_z,-{t_z})$ condition for the summation 
ensures that the Slater determinants in the sum are linearly independent.

The matrix elements of Hamiltonian matrix $\underline V^{\rm IS}$ is given by 
$\underline V^{\rm IS}_{i,j}=\langle\Psi^{\rm IS}_i(S_z)\vert  V^{IS}\vert\Psi^{\rm IS}_k(S_z)\rangle$,
which can be directly determined by (\ref{isoscaLS}). 
One can notice that in the case of the IS pairing the Hamiltonian matrix can be turned also into dyad form
\bea
\underline V^{\rm IS}=-G\frac{1+\vert S_z \vert }{2}{\bf s} {\bf s}^t,
\eea 
where the components of the column vector $\bf s$ are
\bea
&s_i=(-1)^{l{(i)}-{m}{(i)}}(-1)^{\frac{1}{2}-{t_z}{(i)}}.
\eea
Due to the dyad form of the Hamiltonian matrix we know that an eigenvector corresponding to the nonzero eigenvalue 
$-G\frac{1+\vert S_z \vert }{2}{\bf s}^t {\bf s}$ is $\bf s$ and the other eigenvalues are zeros.

This means that  coefficients $B_i$ can be chosen as the components of the normalized eigenvector $\bf s$.
For this choice
\bea\label{Bcoef}
B_i=\frac{(-1)^{l{(i)}-{m}{(i)}}(-1)^{\frac{1}{2}-{t_z{(i)}}}}{\sqrt{\underset{i}{\sum}~1}}
\eea
and the ground state energy of the IS pairing interaction is
\bea
&&-G\frac{1+\vert S_z \vert }{2}
\underset{i}{\sum}s_i^2=-G\frac{1+\vert S_z \vert }{2}\underset{i}{\sum}~1. 
\eea 

From the former formulas, for the $S_z=0$ case of the sd shell we get the wave function
\bea\label{issol}
\frac{1}{\sqrt{12}}
\underset{(n,l,{m},{s_z},{t_z})<(n,l,-{m},-s_z,-{t_z})}{\sum} 
(-1)^{l-m}(-1)^{\frac{1}{2}-{t_z}}
a^{\dagger}_{n,l,{m},{s_z},{t_z}}
a^{\dagger}_{n,l,-{m},-{s_z},-{t_z}}|0\rangle. \nonumber \\
\eea  
This function contains all the 24 modes of the sd shell $\{n,l, m, s_z, t_z\}$ 
and each mode has the same occupation number (same one-mode entropy).
It can be seen that this sp basis is Slater basis, so its elements are natural orbitals,
and the coefficients of the Slater decomposed form are $\lambda_s=\frac{1}{4 \sqrt{3}}$ (see (\ref{sf})). 
The occupation numbers are $4\lambda_s^2=\frac{1}{12}$, that is
the quotient of the particle number and the number of modes that are in the state, 
so (\ref{issol}) has maximum one-body entanglement entropy.

From the proof of \cite{Kruppa21} follows that 
the one-body entanglement entropy and the total correlation possess common upper limit 
so for the considered wave function the total correlation is also maximal.
Since the one-body entanglement entropy is the minimum of the total correlations 
that corresponds to different orthonormal sp bases,
our result implies the interesting observation that for this ground state the total correlation is basis-independent 
and one-mode entropies have the same value in every sp basis.

\section {\label{app3}Associated modes and mutual informations}

For Slater basis we prove that if we fix a mode $i$ and consider all the mode pairs $(i,j)$, 
then the associated mode pair $(i,i_{\rm ass})$ has the largest value of the mutual informations. 
We introduce the notations $x=4\lambda_{s(i)}^2$ and $y=4\lambda_{s(j)}^2$, which are the occupation numbers of the modes $i$ and $j$ that are  present in the wave function.
According to (\ref{mifval}) it is enough to prove
\beq
f(x,y)=h(x)-\big( H(1-x) + H(1-y) - H(1-x-y) \big) \ge 0
\eeq
over the closed  domain
\begin{equation}
D=\{(x,y)\in{\bf{R}}\times{\bf{R}}~|~0\le x\le 1,\quad 0\le y\le 1,\quad 0\le x+y\le 1\}.
\end{equation}

We determine the stationary point of the function $f(x,y)$  with the help of the equations
\bea
&{\partial f(x,y)\over\partial x}=\log_2(1-x-y)-\log_2(x)=0,\nonumber\\
&{\partial f(x,y)\over\partial y}=\log_2(1-x-y)-\log_2(1-y)=0.
\eea
The solution of the system of equations results in the stationary point $(x,y)=(0,1)$, where the function value is zero.
We can conclude that the there is no stationary point in the inner part of $D$, since
the point $(x,y)=(0,1)$ is in the boundary of $D$. 

The domain $D$ is a triangle on the $(x,y)$ plane 
and at the vertex points we have $f(0,0)=0$, $f(1,0)=0$ and $f(0,1)=0$. 
Three line segments are the boundaries of the domain $D$ 
and function $f(x,y)$ has the following forms in these boundaries
\begin{eqnarray}
&f(x,0)=(x-1)\log_2(1-x)-x\log_2(x),\nonumber\\
&f(0,y)=0,\nonumber\\
&f(x,1-x)=0.
\end{eqnarray}
Now we only have to investigate the properties of the function $p(x)=(x-1)\log_2(1-x)-x\log_2(x)$ 
in the interval $[0,1]$. 
The stationary point of $p(x)$ is $x=\frac{1}{2}$ and $p\left(\frac{1}{2}\right)=1$. 
Since $p'(x)=2\frac{{\rm artanh}(1-2x)}{{\rm ln}(2)}$ 
it is easy to see that $p(x)$ is strictly increasing in the interval $[0,\frac{1}{2})$ 
and strictly decreasing on the interval $(\frac{1}{2},1]$. 
All this means that we proved that the function is $f(x,y)$ is nonnegative in the domain $D$.

\end{document}